\documentclass[aps,pra,twocolumn]{revtex4}
\usepackage{graphicx}
\usepackage{rotating}
\usepackage{amsmath}
\usepackage{color}
\usepackage{dcolumn}% Align table columns on decimal point
\usepackage{bm}% bold math

\begin{document}

\title{Spectroscopy of Ultracold, Trapped Cesium Feshbach Molecules}
%==============================================================

\author{M. Mark,$^{1}$ F. Ferlaino,$^{1,2}$ S. Knoop,$^{1}$ J.~G. Danzl,$^{1}$ T. Kraemer,$^{1}$ C. Chin,$^{3}$ H.-C. N\"{a}gerl,$^{1}$ R. Grimm$^{1,4}$}

\affiliation{$^1$Institut f\"{u}r Experimentalphysik and
Forschungszentrum f\"{u}r Quantenphysik, Universit\"{a}t Innsbruck,
6020 Innsbruck, Austria \\
$^2$LENS and Dipartimento di Fisica, Universit\`{a} di Firenze, Firenze, Italy\\
$^3$Physics Department and James Franck Institute, University of
Chicago, Chicago, Illinois 60637\\
$^4$Institut f\"{u}r Quantenoptik und Quanteninformation,
\"{O}sterreichische Akademie der Wissenschaften, 6020 Innsbruck,
Austria }

\date{\today}

\begin{abstract}
We explore the rich internal structure of Cs$_2$ Feshbach
molecules. Pure ultracold molecular samples are prepared in a
CO$_2$-laser trap, and a multitude of weakly bound states is
populated by elaborate magnetic-field ramping techniques. Our
methods use different Feshbach resonances as input ports and
various internal level crossings for controlled state transfer. We
populate higher partial-wave states of up to eight units of
rotational angular momentum ($l$-wave states). We investigate the
molecular structure by measurements of the magnetic moments for
various states. Avoided level crossings between different
molecular states are characterized through the changes in magnetic
moment and by a Landau-Zener tunneling method. Based on microwave
spectroscopy, we present a precise measurement of the
magnetic-field dependent binding energy of the weakly bound
$s$-wave state that is responsible for the large background
scattering length of Cs. This state is of particular interest
because of its quantum-halo character.

%Starting from ultracold Cs atoms in an optical trap we produce up to
%20.000 molecules by magnetic Feshbach association. Following or
%jumping avoided crossings enables to transfer the Cs$_{2}$ dimers
%into different internal molecular states. This allows to analyze the
%underlying molecular energy structure spectroscopically. For our
%spectroscopy measurements we use both magnetic moment and microwave
%techniques. The properties and the coupling strength of different
%molecular states are measured. We show for a $s$-wave molecular
%state the possibility of microwave spectroscopy. We drive a
%bound-bound transition between two molecular states and analyze the
%dependence of the binding energy to the atomic scattering length.
\end{abstract}

\pacs{33.20.-t, 33.80.Ps, 34.50.-s, 05.30.Jp}
%\keywords{}

\maketitle


%==============================================================
\section{Introduction\label{Introduction}}
%==============================================================

%----allgemein das feld------

The possibility to associate molecules via Feshbach resonances in
ultracold gases \cite{Koehler2006} has opened up new avenues of
research. The demonstration of coherent atom-molecule coupling
\cite{Donley2002amc}, the creation of pure molecular samples from
atomic Bose-Einstein condensates
\cite{Herbig2003,Duerr2004mols,Xu2003}, and the formation of
ultracold molecules from atomic Fermi gases
\cite{Regal2003cum,Strecker2003coa,Cubizolles2003pol,Jochim2003pgo}
paved the way for spectacular achievements. Prominent examples are
the observation of molecular Bose-Einstein condensation
\cite{Jochim2003, Greiner2003, Zwierlein2003} and the creation of
strongly interacting superfluids in atomic Fermi gases
\cite{Inguscio2006ufg}. Ultracold molecules have also opened up
new ways to study few-body physics with ultracold atoms
\cite{Chin2005,Kraemer2006efe}. In optical lattices, controlled
molecule formation
\cite{Stoferle2006mof,Thalhammer2006llf,Ospelkaus2006uhm} has been
the experimental key to create novel correlated states in a
crystal-like environment \cite{Winkler2006rba,Volz2006pqs}.

%----Feshbach association------
%In all these experiments, the control of the interatomic interaction
%by magnetic fields plays an essential role in the association
%process.
A Feshbach resonance \cite{Tiesinga1993,Inouye1998} arises when a
bound molecular dimer state is magnetically tuned near a two-atom
scattering state, leading to resonant atom-molecule coupling. The
molecular structure and in particular the molecular state that
interacts with the atomic threshold determine the character of a
particular Feshbach resonance \cite{Koehler2006}. The rotational
angular momentum of the molecular state, characterized by the
rotational quantum number $\ell$, plays a central role. Various
types of Feshbach molecules ranging from dimers in $s$-wave states
($\ell=0$) to dimers in $g$-wave states ($\ell=4$) have been
realized \cite{Koehler2006}.
%\cite{Xu2003,Mukaiyama2004,Duerr2004mols,Volz2005,Herbig2003}.
%A recent review on Feshbach molecules can be found
%in\,\cite{Koehler2006}.

%----Cs and its special role----
For experiments with molecular quantum gases, cesium is particularly
rich as it offers a unique variety of different Feshbach resonances
and molecular states \cite{Chin2004}. %The complexity of the Cs energy structure is
%connected with near-resonant scattering properties at low magnetic
%fields and originates from
Pronounced relativistic effects lead to strong higher-order coupling
between atom pairs and molecules and between different molecular
states. For achieving Bose-Einstein condensation in cesium
\cite{Weber2003}, the detailed understanding of the complex
molecular structure was a crucial factor. The interaction properties
of cesium atoms were characterized by Feshbach spectroscopy in a
series of atom scattering experiments performed
%by Chin, Kerman, Vuletic, and Chu
at Stanford University \cite{Vuletic1999,Chin2000,Chin2004}. In
these experiments the magnetic field positions of many Feshbach
resonances up to $g$-wave character were measured. This provided the
necessary experimental input for theoretical calculations of the
molecular energy structure \cite{Leo2000,Chin2004}, performed at the
National Institute of Standards and Technology (NIST). In the
following, we will refer to the cesium molecular structure as
presented in Ref.~\cite{Chin2004} as the ``NIST model''. It
represents the current knowledge of the structure of weakly bound
molecular states, and thus constitutes the theoretical basis for the
experiments discussed in this work.

%----was hebt dieses paper ab von anderen----
In this Article, we report on a thorough investigation of the
energy structure of weakly bound Cs$_2$ Feshbach molecules. Our
experiments are performed on ultracold molecular samples confined
in a CO$_2$-laser trap
\cite{Takekoshi1998,Chin2005,Staanum2006eio,Zahzam2006amc}
%In Ref.~, the structure of weakly bound molecular states was derived
%from the observation of Feshbach resonances in atom-atom scattering
%experiments with the help of the theoretical NIST model.
and extend previous work \cite{Chin2004} in three important ways.
First, we show how any of the weakly bound molecular states can be
populated based on elaborate time-dependent magnetic field
control. Spectroscopy performed on various molecular states
confirms the main predictions of the NIST model and provides input
for further refinements of the model. Second, we demonstrate how
one can indirectly populate states with high rotational angular
momentum of $\ell=8$ ($l$-wave states) by taking advantage of
avoided level crossings with $\ell=4$ ($g$-wave) states. For these
$l$-wave states, direct Feshbach association is not feasible
because of negligible coupling with the atomic scattering
continuum. Third, spectroscopy on avoided crossings between bound
states yields precise information about the coupling strengths
between molecular states.

%----summary/inhalt----
In Sec.\,\ref{Sec_specialCs}, we first review the energy structure
of weakly bound Cs$_2$ dimers. In Sec.\,\ref{preparation}, we
address the preparation of molecular samples, detail our
techniques to transfer molecular samples to various internal
states, and present the methods for molecule detection. In
Sec.\,\ref{spectroscopy}, we report on spectroscopic measurements
using magnetic moment and microwave techniques.

%Weak avoided crossings allow for ramsey interferometry between
%different molecular states.

%=============================================================================
\section{Energy structure of weakly bound Cesium dimers}\label{Sec_specialCs}
%=============================================================================

%Weakly bound Cs dimers exhibit a highly complex energy structure.
%Current knowledge of the molecular spectrum near threshold is based
%on theoretical calculations of the molecular energy structure at the
%National Institute of Standards and Technology (NIST) in combination
%with Feshbach loss spectroscopy of ultracold Cs atoms performed at
%Stanford University\, \cite{Chin2004}. In the following, we will
%refer to the cesium molecular structure as presented in
%Ref.\,\cite{Chin2004} as the ``NIST model''.

\begin{figure}
\includegraphics{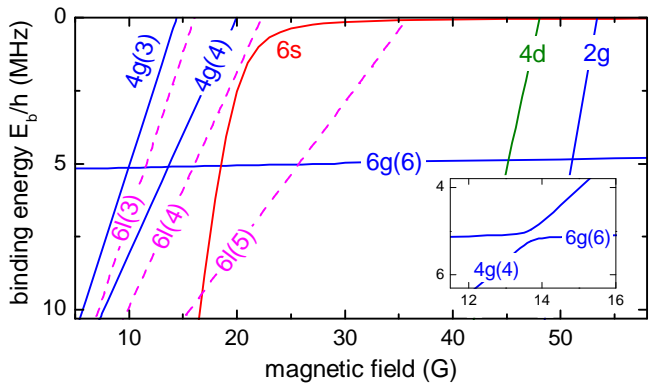}
\caption{(color online) Molecular energy structure below the
threshold of two free Cs atoms in the absolute ground state $\vert
F\!=\!3, m_{F}\!=\!3\rangle$. Molecular state labeling is
according to the quantum numbers $f \ell (m_f)$,
$m_{\ell}=6-m_{f}$. The quantum number $m_{f}$ is omitted for
states with $m_{f}\!=\!f$ and $m_{\ell}\!=\!\ell$. The solid lines
represent the $s,d$ and $g$-wave states included in the NIST
model\,\cite{Chin2004}. The intersections of the $d$- and $g$-wave
states with the threshold cause narrow Feshbach resonances that
can be used for molecule production. The curvature of the $6s$
state arises from a large avoided crossing between two states of
the same $f \ell$ quantum numbers. The NIST model does not take
into account weak avoided crossings between bound molecular states
mediated by the relativistic spin-spin dipole and second order
spin-orbit interactions. If these interactions are taken into
account, the crossings between bound molecular states become
avoided as illustrated in the inset for the example of the
$4g(4)/6g(6)$ crossing. The dashed lines represent $l$-wave states
($\ell=8$) obtained from extended calculations based on the NIST
model.\label{fig_roadmap}}
\end{figure}

Figure\,\ref{fig_roadmap} gives an overview of the molecular
states relevant to the present work, covering the magnetic field
region up to 55\,G and binding energies up to $h\times 10$\,MHz,
where $h$ is Planck's constant. Zero energy corresponds to the
dissociation threshold into two Cs atoms in the absolute hyperfine
ground state sublevel $\vert F\!=\!3, m_{F}\!=\!3\rangle$. Each
intersection of the atomic threshold with a molecular state
corresponds to a Feshbach resonance. The rotational angular
momentum associated with a molecular state is denoted by the
quantum number $\ell$. We follow the convention of labeling states
with $\ell=0,2,4,6,8,\ldots$ as $s,d,g,i,l,\ldots$ -wave
states\,\cite{Russell1929} and the associated Feshbach resonances
as $s,d,g,i,l,\ldots$ -wave resonances. As a consequence of the
bosonic nature of Cs atoms, only even values of $\ell$ occur. The
solid lines in Fig.\,\ref{fig_roadmap} represent states resulting
from the NIST model\,\cite{Chin2004}, including $s$-, $d$- and
$g$-wave states. For two interacting Cs atoms, relativistic
spin-spin dipole and second-order spin-orbit interactions are
particularly important \cite{Leo2000}. Therefore, in Cs not only
$s$- and $d$-wave states but also $g$-wave states couple
sufficiently to the atomic threshold to produce experimentally
observable Feshbach resonances. The magnetic field positions of
these Feshbach resonances were determined experimentally in an
optically confined atomic Cs
gas\,\cite{Vuletic1999,Chin2000,Chin2004}. The NIST predictions
for the weakly bound molecular structure result from a theoretical
model of the energy spectrum with parameters adjusted to reproduce
the measured magnetic field positions of the Feshbach resonances.

Cs molecular states near threshold are for the most part
sufficiently well characterized by the quantum numbers $|f$,
$m_{f}$; $\ell$, $m_{\ell}\rangle$\,\cite{Koehler2006}, where $f$
represents the sum of the total atomic spins $F_{1,2}$ of the
individual atoms, and $\ell$ is the nuclear mechanical angular
momentum quantum number. The respective projection quantum numbers
are given by $m_{f}$ and $m_{\ell}$. In special cases the quantum
numbers $F_{1}$ and $F_{2}$ also have to be specified. To account
for the molecular structure below threshold, not only the exchange
and van der Waals interaction, the atomic hyperfine structure, and
the Zeeman energy, but also the weaker relativistic spin-spin
dipole and second-order spin-orbit interactions have to be
considered\,\cite{Mies1996, Chin2004}. The exchange and van der
Waals interactions conserve $\ell$ and $f$, whereas the two
relativistic interactions weakly mix states with different $\ell$
and $f$. The complete interaction Hamiltonian conserves the total
angular momentum $f+\ell$ at zero magnetic field. More
importantly, it always conserves the projection of the total
angular momentum $m_{f}+m_{\ell}$. In our experiments, we start
with an ultracold, spin-polarized atomic sample of Cs atoms in
their hyperfine ground state $\vert F\!=\!3, m_{F}\!=\!3\rangle$.
At ultralow scattering energies only incoming $s$-waves
($\ell\!=\!0$) need to be considered. The atomic scattering state
is hence $\vert f\!=\!6, m_{f}\!=\!6; \ell\!=\!0,
m_{\ell}\!=\!0\rangle$. Consequently all molecular states relevant
to the present work obey $m_{f}+m_{\ell}=6$.

To label molecular states we use the three quantum numbers $f
\ell(m_{f})$. For states with $m_{f}=f$ and $m_{\ell}=\ell$, we
only use $f$ and $\ell$ for brevity.
Table\,\ref{tab_quantumnumbers} gives the full set of angular
momentum quantum numbers for all molecular states relevant to the
present work.

\begin{table}
\caption{List of angular momentum quantum numbers for the relevant
molecular states. Each state is represented by four quantum
numbers: the total internal angular momentum $f$ and the
rotational angular momentum $\ell$ with $m_{f}$ and $m_{\ell}$ as
the respective projections along the quantization axis.
\label{tab_quantumnumbers}}
\begin{ruledtabular}
\begin{tabular}{c c c c c c c c c c}

   label of & 6$s$ & 4$d$ & 2$g$ & 4$g$(3) & 4$g$(4) & 6$g$(6) & 6$l$(3) & 6$l$(4) & 6$l$(5) \\
   state\\
  \hline
  $f, m_{f}       $  & $6$,$6$ & $4$,$4$ & $2$,$2$ & $4$,$3$ & $4$,$4$ & $6$,$6$ & $6$,$3$ & $6$,$4$ & $6$,$5$ \\
  $\ell, m_{\ell} $  & $0$,$0$ & $2$,$2$ & $4$,$4$ & $4$,$3$ & $4$,$2$ & $4$,$0$ & $8$,$3$ & $8$,$2$ & $8$,$1$ \\
  \end{tabular}
\end{ruledtabular}
\end{table}

Coupling between molecular states with the same $f$ and $\ell$ in
general leads to very broad avoided crossings between molecular
states. The strong curvature of the $6s$ state in
Fig.\,\ref{fig_roadmap} is a result of such a crossing. In this
case, a weakly bound 6s-state with $F_{1}\!=\!3$ and $F_{2}\!=\!3$
happens to couple to a 6s-state with $F_{1}\!=\!4$ and
$F_{2}\!=\!4$. Narrow avoided crossings arise when molecular
states of different $f$ and $\ell$ intersect. These narrow
crossings are mediated by the spin-spin dipole and second-order
spin-orbit interactions. In the NIST model narrow avoided
crossings were only taken into account for special cases where it
was necessary to assign the experimentally observed Feshbach
resonances. Consequently, the molecular states in
Fig.\,\ref{fig_roadmap} are shown as intersecting lines.
Nevertheless, the existence of avoided crossings between molecular
states of different $f\ell$ is crucial for the present work as it
allows the transfer of molecules from one state to another. As an
example, the inset in Fig.\,\ref{fig_roadmap} schematically
illustrates the avoided crossing between the $4g(4)$ state and the
$6g(6)$ state at $\sim 13.5$\,G.

The dashed lines in Fig.\,\ref{fig_roadmap} represent $l$-wave
states. As states with higher angular momentum ($\ell > 4$) do not
couple to the $s$-wave scattering continuum, the $l$-wave states
cannot be observed by Feshbach spectroscopy in an ultracold atomic
gas. Consequently, no experimental input for higher angular
momentum states was available for the NIST model. It is not a
surprise, however, that for Cs $l$-wave states exist in the low
magnetic field region. This follows from a general property of the
asymptotic van der Waals potential \cite{Gao2000}. In the case of
an $s$-state being close to threshold, angular momentum states
with $\ell=4, 8, \ldots$ should also occur near threshold. The
observation of both $g$- and $l$-wave states in a system with
near-resonant s-wave background scattering properties nicely
illustrates this general property. When the NIST model is extended
to states with higher angular momentum it indeed predicts $l$-wave
states in the low-field region \cite{TiesingaPrivate}. The
calculations are expected to accurately predict the magnetic
moments, i.\,e.\,the slopes, of these states but they leave some
uncertainty concerning the exact binding energies
\cite{TiesingaPrivate}. The $l$-wave states shown in
Fig.\,\ref{fig_roadmap} result from the extended NIST model, but
they are energetically adjusted to the experimental observations
(Sec.~\ref{spectroscopy}) by equally down-shifting all three
states by about $h \times 2$\,MHz.

%===================================================================================
\section{Preparation of C$\textbf{\textrm{s}}$$_2$
Molecules in various internal states}\label{preparation}
%===================================================================================

In this Section, we present our basic methods to prepare Cs$_2$
Feshbach molecules in various internal states. The starting point
for all the experiments is an optically trapped ensemble of Cs
atoms, the preparation of which is briefly summarized in
Sec.~\ref{prep_at}. We then describe the creation of optically
trapped Cs molecules based on different Feshbach resonances
(Sec.~\ref{creation}). These resonances serve as ``entrance
doors'' into the rich molecular structure near threshold. In
Sec.~\ref{state_transfer}, we discuss our techniques to transfer
molecules to various internal states by application of elaborate
time-variations of the magnetic field. We make use of the
possibility of adiabatic or diabatic passages through avoided
crossings. In Sec.~\ref{detection} we discuss the methods to
detect the molecular samples through controlled dissociation.

%-----------------------------------------------------------------------------------
\subsection{Atomic sample preparation}
\label{prep_at}
%-----------------------------------------------------------------------------------

The setup used for the present experiments is optimized for
molecule trapping and molecular state manipulation, and not for
Bose-Einstein condensation (BEC) as in our previous work
\cite{Weber2003,Kraemer2004,Herbig2003}. Here we start with an
atomic sample near degeneracy, for which we obtain sufficient
efficiencies for molecule formation.

\begin{figure}
\includegraphics[width=8.5cm]{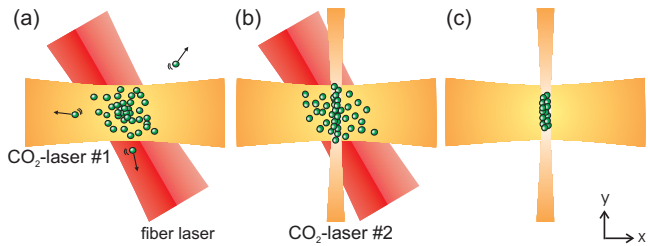}
\caption{(color online) Successive stages of optical dipole traps.
(a) We first realize a large volume ``reservoir'' trap by crossing
a CO${_2}$-laser beam and a 1070nm fiber-laser beam in the
presence of a levitating gradient. (b) We ramp up a tightly
focused CO$_2$-laser beam, (c) switch off the 1070\,nm reservoir
beam and then evaporate along the vertical direction (z-axis) by
lowering the gradient. We obtain typically $4\times 10^5$ Cs atoms
at a temperature of $200$\,nK. \label{figTrapscheme}}
\end{figure}

For the present experiments we use a sequence of three dipole
traps in the cooling and sample preparation process as shown in
Fig.\,\ref{figTrapscheme}. The final dipole trap for molecule
experiments is realized by crossing two CO$_2$-laser beams. The
far-infrared CO$_2$-laser trap avoids the use of near-infrared
radiation. In previous experiments we used the 1064-nm broad-band
radiation from an Yb fiber-laser in the final trapping stage, and
we observed strong light-induced trap losses for the Feshbach
molecules, presumably as a result of excitation of molecular
bound-bound transitions. The CO$_2$-light is sufficiently off
resonance and it thus allows for long molecule trapping times
\cite{Chin2005,Staanum2006eio,Zahzam2006amc} and  facilitates
efficient in-trap production of molecules. One of the important
features in our previous experiments on Cs BEC and the production
of Feshbach molecules is the ability to levitate the atoms and
molecules against gravity using a magnetic field gradient
\cite{Weber2003,Chin2005}. However, for the preparation of
molecular samples in various states the requirement of magnetic
field gradients is problematic, because molecules can have widely
different magnetic moments and thus require different levitation
gradients. By using a relatively tight focus of one of the
trapping CO$_2$-laser beams, we can hold the molecules against
gravity without the levitation gradient field.

The cooling and trapping procedure for the atoms is similar to the
techniques described in Ref.~\cite{Kraemer2004}. In brief, we
first load a magneto-optical trap (MOT) followed by a short
optical molasses phase to compress and further cool the atomic
sample. Using the technique of Raman sideband cooling in an
optical lattice\,\cite{Treutlein2001} the atoms are then cooled
and simultaneously polarized into the lowest hyperfine state
$|F\!=\!3,m_F\!=\!3\rangle$. We typically obtain $2 \times 10^{7}$
atoms at a temperature of $\sim 700$\,nK.

The polarized sample is adiabatically released from the lattice
into a large volume dipole trap. This ``reservoir trap'' is
realized by two crossed laser beams. As illustrated in
Fig.\,\ref{figTrapscheme}(a), we use a CO$_{2}$-laser and an Yb
fiber-laser for the reservoir with wavelengths of $10.6$\,$\mu$m
and $1070$\,nm, respectively. For each laser the beam waist is
around $650$\,$\mu$m. This shallow reservoir trap cannot hold the
atoms against gravity. Therefore we apply magnetic levitation at
this stage \cite{Weber2003,Kraemer2004}. The resulting effective
trap depth is about $7$\,$\mu$K. After releasing the atoms from
the optical lattice used for Raman sideband cooling into the
reservoir trap, $2$\,s of plain evaporation are necessary to
thermalize the sample in the trap. The thermalization is performed
at a magnetic field of $75$\,G, corresponding to a scattering
length of about $1200$\,a$_0$, where a$_0$ denotes Bohr's radius.
We measure about $4\times 10^{6}$ atoms at a temperature of $\sim
\!1\mu$K\,\cite{Kraemer2004}, the phase-space density is $\sim
1/1000$.

After thermalization, the reservoir trap is crossed with a tightly
focused CO$_{2}$-laser beam as shown in
Fig.\,\ref{figTrapscheme}(b). The waist of this ``CO$_2$ dimple''
is about $80$\,$\mu$m. This value is diffraction-limited by the
aperture of the window of the vacuum chamber. We linearly ramp up
the power of the beam within $2.8$\,s to $\sim 2.5$\,W
corresponding to a trap depth of about $17$\,$\mu$K.
Simultaneously the magnetic field is ramped down to $35$\,G,
corresponding to a scattering length of $700$\,a$_0$. This
procedure provides efficient collisional loading of the CO$_2$
dimple \cite{Kraemer2004}. The remaining atoms in the reservoir
trap are released by switching off the Yb fiber-laser beam. In the
crossed CO$_2$-laser trap (see Fig.\,\ref{figTrapscheme}(c)) we
measure typically $1\times 10^{6}$ atoms at a temperature of
$1$\,$\mu$K.

We then apply forced evaporative cooling by exponentially lowering
the magnetic field gradient within $6.3$\,s to zero. Atoms thus
mainly escape from the trap along the vertical direction.
Simultaneously we adjust the scattering length by decreasing the
magnetic field to a final value of $\sim 22$\,G to minimize
three-body losses \cite{Weber2003a}. As we also slightly decrease
the power of the CO$_2$ dimple to $2$\,W, the effective trap depth
without levitation is $\sim 2$\,$\mu$K.

We finally obtain $\sim 4\times 10^{5}$ Cs atoms at a temperature
of about $200$\,nK in the crossed CO$_2$-laser trap. The trap
frequencies of the final configuration without magnetic field
gradient are measured to be 84(1)\,Hz and 10(1)\,Hz in the
horizontal plane, and 80(1)\,Hz in the vertical direction. The
resulting peak density of the atoms is $\sim$$1 \times
10^{13}$\,cm$^{-3}$, and the phase-space density is about $0.4$.

%-----------------------------------------------------------------------------------
\subsection{Molecule production through Feshbach resonances}\label{creation}
%-----------------------------------------------------------------------------------

\begin{figure}
\includegraphics[width=8.5cm]{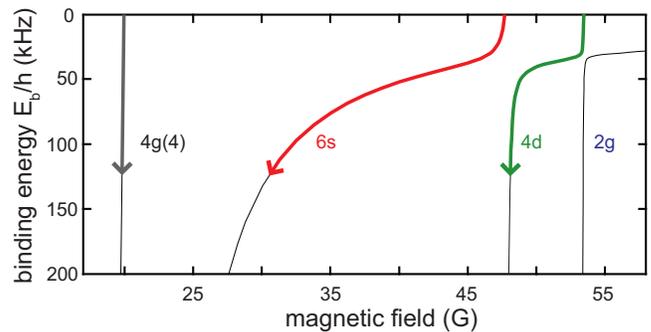}
\caption{(color online) The molecular energy structure for very
small binding energies in the region of the $19.8$\,G, $47.9$\,G
and $53.4$\,G Feshbach resonances, not resolved in
Fig.\,\ref{fig_roadmap}. Above $45$\,G two avoided crossings are
present in the energy structure. We use all three Feshbach
resonances to associate molecules by ramping or switching the
magnetic field. The arrows indicate the pathway after molecule
association as the magnetic field is ramped down to allow for the
optical removal of the atoms from the molecules. For details see
text. \label{fig48Gstates}}
\end{figure}

We magnetically associate ultracold cesium atoms to dimers on
Feshbach resonances
\cite{Herbig2003,Xu2003,Duerr2004mols,Koehler2006}. In this work
we use \emph{three different} resonances, the two $g$-wave
resonances at $B=19.8$\,G and $53.4$\,G and the $d$-wave resonance
at $47.9$\,G, see Fig.\,\ref{fig48Gstates}. The width of the
$g$-wave resonances is only a few mG, the $d$-wave resonance is
about $200$\,mG wide. While the molecule formation at the
$19.8$\,G $g$-wave resonance results in $g$-wave molecules, the
association at the $47.9$\,G $d$-wave resonance leads in practice
to $s$-wave molecules. This is a consequence of an avoided level
crossing close to threshold between the states 4$d$ and 6$s$, see
Fig.\,\ref{fig48Gstates}. Similarly, the association at the
$53.4$\,G $g$-wave resonance results in $d$-wave molecules.

For molecule creation two different techniques are used. Depending
on the character of the Feshbach resonance, we apply a ramping or
a switching scheme to produce dimers \cite{Mark2005}. The
switching scheme works particularly well at the narrow $g$-wave
Feshbach resonances. We set the magnetic field typically $0.5$\,G
above the resonance. The field is then suddenly changed to the
resonance position and kept there for $\sim 5$\,ms. In contrast,
at the much broader $47.9$\,G $d$-wave Feshbach resonance we find
superior efficiency by applying a linear magnetic field ramp
(ramping scheme). We start typically $100$\,mG above the resonance
and linearly ramp the magnetic field within $5$\,ms to about
$100$\,mG below the resonance. The efficiencies for molecule
production range from a few percent up to 20\%. Starting from
$4\times 10^5$ atoms we typically obtain $15,000$ molecules, see
Table~\ref{tab_moleculeproduction}.

To prepare a maximum number of molecules in the trap, it is
necessary to separate atoms and molecules as fast as possible,
since atom-dimer collisions dramatically reduce the lifetime of
the molecular sample \cite{Mukaiyama2004}. We remove the atoms
from the dipole trap using a `blast' technique similar to
Ref.~\cite{Xu2003}. First, the atoms are pumped out of the
$|3,3\rangle$ state by light close to the $F=3 \rightarrow
F'\!=\!3$ transition. The blast pulse is tuned to the closed
optical transition $|F\!=\!4,m_{F}\!=\!4\rangle \rightarrow |F'=5,
m_{F'}\!=\!5\rangle$, which we also use for imaging. The optical
cleaning process causes some unwanted loss and heating of the
molecules. Particularly if the molecules are very weakly bound
($\lesssim h\times 1$\,MHz) or the blast duration is too long
($\gtrsim 1$\,ms) these effects are not negligible. Therefore
immediately after the association we rapidly ramp the magnetic
field further down. When the binding energy of the molecular state
is on the order of $h \times 5$\,MHz the molecules are much less
affected by the blast light. When using the $19.8$\,G resonance
the magnetic field has to be ramped down only a few Gauss to reach
such a binding energy. In case of the $47.9$\,G resonance
($53.4$\,G resonance) the lowering of the magnetic field transfers
the molecules into the 6$s$-state (4$d$-state) through the present
avoided crossings, see Fig.\,\ref{fig48Gstates}. Therefore, to
reach a sufficiently large binding energy a larger change in the
magnetic field is required, resulting in a longer time to reach
the field. However, with a typical blast duration of $400$\,$\mu$s
we achieve a sufficient removal of the atoms from the trap while
keeping the blast-induced molecule losses small ($\sim 10-15$\%).
%In fact, the blast pulse timing is
%limited by the finite response time of the magnetic field, mainly
%resulting from eddy currents, for details see also appendix.

\begin{table}
\caption{Parameters for molecule production using three different
Feshbach resonances. $B_r$ is the field value at which the atoms
are removed with the blast technique.
\label{tab_moleculeproduction}}
\begin{ruledtabular}
\begin{tabular}{l c c c}
  Feshbach resonance position (G)  & $19.8$ & $47.9$ & $53.4$ \\
   \hline
  entrance state            & 4$g$(4) & 4$d$ & 2$g$ \\
  ramp speed (G/s)          & - & 36 &  - \\
  state at $B_r$            & 4$g$(4) & 6$s$ & 4$d$ \\
  $B_r$ (G)                 & 14.5 & 19.7 & 45\\
  time to reach $B_r$ (ms) & $0.4$ & $3$ & $1$ \\
  number of molecules & $19000$ & $9000$ & $15000$ \\
  \end{tabular}
\end{ruledtabular}
\end{table}

The precise timing for molecule production, the magnetic field for
the purification, and the obtained number of molecules strongly
depend on the particular Feshbach resonance.
Table\,\ref{tab_moleculeproduction} summarizes the relevant
experimental parameters of our molecule production. We measure a
typical temperature of $250$\,nK for the molecular samples. This
is slightly higher than the temperature of the atoms, presumably
because of the effects of the blast cleaning technique. The
corresponding peak density of the molecules is $\sim 7 \times
10^{11}$\,cm$^{-3}$.

%-----------------------------------------------------------------------------------
\subsection{Molecular state transfer}\label{state_transfer}
%-----------------------------------------------------------------------------------
Other molecular states than the ones that we can directly access
through the Feshbach creation schemes can be populated by
controlled state transfer. The experimental key is the precise
control of Landau-Zener tunneling at avoided crossings through
elaborate magnetic field ramps. By means of the ramp speed we can
choose whether a crossing is followed adiabatically (slow ramp) or
jumped diabatically (fast ramp). An important application of
controlled ramps through avoided crossings is the coherent
splitting of the molecular wave function for intermediate ramp
speeds, as reported in Ref.~\cite{Mark2007}.

\begin{figure*}
\includegraphics[width=17cm]{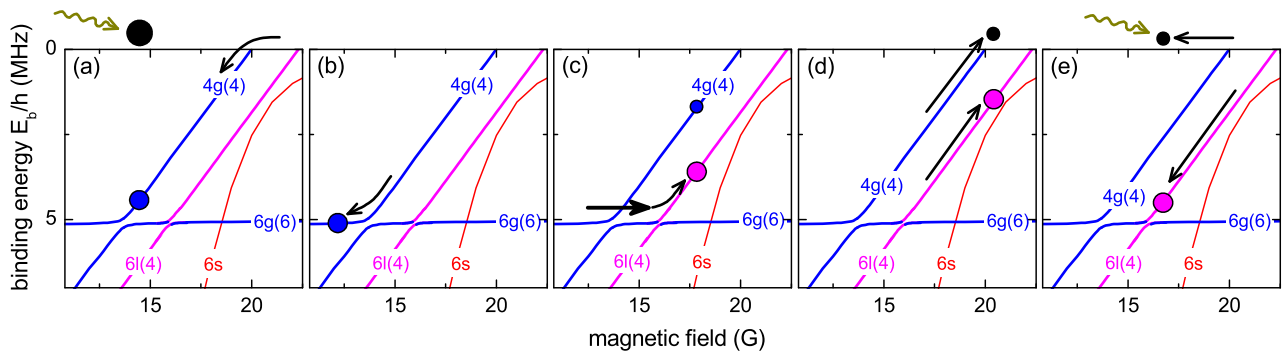}
\caption{(color online) Illustration of the experimental toolbox
for the preparation of molecules in various internal states. As an
example the creation of molecules in the 6$l$(4) state is shown.
(a) First, we produce 4$g$(4) molecules at the $g$-wave Feshbach
resonance at $19.8$\,G and remove the remaining atoms by  a short
blast pulse, indicated by the rippled arrow. (b) The 4$g$(4) state
is transferred into state 6$g$(6) by slowly lowering the magnetic
field. (c) To overcome the avoided level crossing between state
4$g$(4) and 6$g$(6) a very fast magnetic field ramp is applied.
The target state 6$l$(4) is then accessed by using a second
adiabatic ramp. (d) To remove the residual 4$g$(4) molecules from
the 6$l$(4) dimers we ramp the magnetic field slightly above
$20$\,G, dissociating the 4$g$(4)-molecules into atoms while not
affecting the $l$-wave molecules. (e) The magnetic field is
lowered again to increase the binding energy of the target state
molecules. A second blast pulse removes the remaining atoms.
\label{figStateJumping}}
\end{figure*}

Within the Landau-Zener model \cite{Landau1932xxx,Zener1932xxx} an
avoided level crossing is characterized by two parameters, the
coupling strength and the differential slope of the states. For
the coupling strength we introduce the parameter $V$ as half the
energy splitting between the two states at the crossing point. To
characterize the slope we use $\Delta \mu$ as the magnetic moment
difference between the two states. With these two parameters one
commonly defines a critical ramp speed
\begin{equation}\label{eq_Bcrit}
r_{c}=\frac{2 \pi V^{2}}{\hbar \Delta \mu}.
\end{equation}
For fast ramps with ramp speed $\dot{B} \gg r_{c}$, the passage
through the crossing is diabatic and the molecules stay in the
same bare state. For slow ramps ($\dot{B} \ll r_{c}$), an
adiabatic transfer into the other molecular state takes place. For
Cs Feshbach dimers the typical coupling strengths for crossings
between states of different $f \ell$ (see
Sec.~\ref{Sec_specialCs}) are such that the critical ramps speeds
are found in a range convenient for experiments. Full control
ranging from completely diabatic Landau-Zener tunneling to full
adiabatic transfer can be achieved for most crossings (see
Sec.\,\ref{sec_avoidedcross}).

%-------------------THE BIG EXAMPLE --------------------
To illustrate the experimental procedure for transferring
molecules into different states, we now consider the preparation
of a molecular sample in a selected ``target'' state. As an
example we discuss the population of the target state 6$l$(4) in
detail. As the state transfer strongly relies on the technical
performance of the set-up for magnetic field control, we give a
detailed description of the set-up in the Appendix~\ref{appendix}.

%-------------End intro target state ----------

As shown in Fig.~\ref{figStateJumping}(a), we first create 4$g$(4)
molecules at the $19.8$\,G Feshbach resonance. We then lower the
magnetic field to about $14.5$\,G and remove the remaining atoms
with the blast pulse. In a second step, see
Fig.~\ref{figStateJumping}(b), we lower the magnetic field to
$\sim 12$\,G within a few ms. Consequently, we pass the avoided
crossing between the two states 4$g$(4) and 6$g$(6) at about
$13.3$\,G. For this crossing the critical ramp speed, given by
Eq.\,\ref{eq_Bcrit}, is $r_c \sim 1100$\,G/ms as the coupling
strength is $V \simeq h \times 150$\,kHz \cite{Chin2005}. With the
applied ramp speed of $\sim 2$\,G/ms the transfer into state
6$g$(6) is therefore fully adiabatic.
Fig.\,\ref{figStateJumping}(c) illustrates the transfer of the
6$g$(6) molecules to the target state 6$l$(4). First we apply a
fast magnetic field ramp to overcome the 4$g$(4)/6$g$(6) crossing,
indicated by the straight arrow. The high ramp speed required is
accomplished by a specially designed ``booster'' coil, described
in the Appendix. With a maximum possible ramp speed of
$7500$\,G/ms we achieve a transfer efficiency of typically $70$\%.
After the jump we enter the target state 6$l$(4) by adiabatically
following the next avoided crossing between state 6$g$(6) and
6$l$(4) at $\sim 15.5$\,G. For this crossing we find a fully
adiabatic transfer when ramping the magnetic field from $15$\,G to
$\sim 17$\,G within a few ms. In the fourth step, illustrated in
Fig.\,\ref{figStateJumping}(d), we prepare the cleaning of the
sample from the residual 4$g$(4) molecules. The magnetic field is
ramped up to $\sim 20$\,G and kept constant for a few ms. While
the remaining 4$g$(4) molecules break up into atoms, the 6$l$(4)
molecules are not affected as their dissociation threshold is
higher. Finally, we ramp down the magnetic field to $B\simeq16$\,G
where the target molecules are well below threshold, see
Fig.\,\ref{figStateJumping}(e). Again we remove the residual atoms
using a blast pulse. As a result, we obtain a pure molecular
sample in the state 6$l$(4).

In analogous ways, we apply these techniques to populate any of
the states shown in Fig.\,\ref{fig_roadmap}.

%---------------------------------------------------------------------------------
\subsection{Molecule detection}\label{detection}
%---------------------------------------------------------------------------------

%start with a general idea.
%------------------------------
The standard detection scheme for Feshbach molecules relies on the
controlled dissociation by reverse magnetic field
ramps\,\cite{Herbig2003,Duerr2004}. When ramping the magnetic field
above the dissociation theshold, the molecules become quasi-bound
and decay into the atomic scattering continuum. The resulting atom
cloud can then be detected using standard absorption imaging.

%output ports
%------------
%As we use three different Feshbach resonances to associate $s$-,
%$d$- or $g$-wave molecules,

Magnetic dissociation by inverse magnetic field ramps is
straightforward for states with large coupling to the scattering
continuum, and hence any of the Feshbach resonances up to $g$-wave
can be used. We ramp the magnetic field typically $2$\,G above
threshold and wait a few ms at the dissociation field before the
image is taken. $l$-wave molecules do not sufficiently couple to
the atomic continuum and significant dissociation is prevented.
One way to detect $l$-wave dimers is to transfer these molecules
into one of the $s$-, $d$- or $g$-wave states which allow for
dissociation and hence for detection.

\begin{figure}
\includegraphics[width=8.0cm]{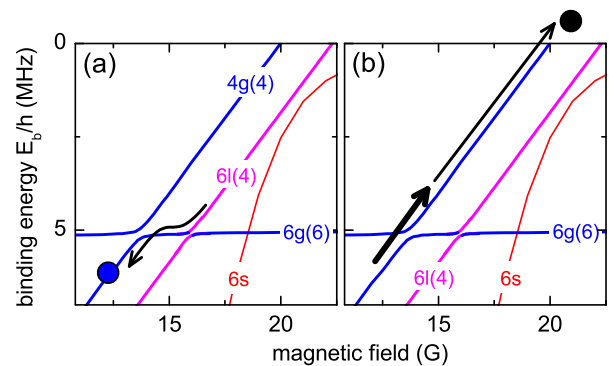}
\caption{(color online) Example of a detection scheme for $l$-wave
molecules. (a) First the 6$l$(4) molecules are adiabatically
transferred into the state 4$g$(4) via the state 6$g$(6) by
ramping down the magnetic field. (b) The avoided crossing at
$13.6$\,G is passed by using a fast magnetic field ramp. When
ramping up to a field of $\sim 21$\,G, the molecules are brought
above threshold and dissociate. The resulting atom cloud is
detected using the standard absorption imaging
technique.\label{fig_DetectionScheme}}
\end{figure}

To illustrate the detection by controlled dissociation, we resume
our previous example of Sec.\,\ref{state_transfer}, where we have
described the preparation of a molecular sample in the 6$l$(4)
state. Fig.\,\ref{fig_DetectionScheme} shows the detection scheme
that we use for this state. First the molecular sample is
adiabatically transferred to the state 4$g$(4) via the state
6$g$(6) by lowering the magnetic field to about $12.5$\,G, see
Fig.~\ref{fig_DetectionScheme}(a). We then perform a diabatic
state transfer over the avoided crossing at $\sim$14\,G as
indicated in Fig.~\ref{fig_DetectionScheme}(b). Finally, we ramp
the magnetic field up to $\sim$21\,G, which is well above the
dissociation threshold of the 4$g$(4) state.

%how to detect l-wave molecules
%------------------------------
An alternative method for the detection of $l$-wave molecules
relies on the particular energy structure of Cs atoms above the
dissociation threshold. We find that the decay of metastable
$l$-wave dimers can be mediated by coupling to a quasi-bound
$g$-wave molecular state above threshold. Such a coupling with
$\Delta \ell=4$ is sufficiently strong. We have previously used
this process for the detection of $l$-wave molecules in the state
6$l$(3) in Ref.~\cite{Mark2007}. A more detailed analysis of this
dissociation mechanism will be presented elsewhere
\cite{Knoop2007}.

%what light do we use
%--------------------
For imaging of the resulting atomic cloud, the atoms are first
pumped to the $|4,4\rangle$ state using light close to the
$F\!=\!3 \rightarrow F'\!=\!3$ transition. The imaging light is
resonantly tuned to the closed $|F\!=\!4, m_F\!=\!4\rangle
\rightarrow |F'\!=\!5,m_F'\!=\!5\rangle$ optical transition,
taking the Zeeman shift at the imaging magnetic field into
account.

%rudiiiiiiiiiiiiiiiiiiiiiiiiiiiiiiiiiiiiiiiiiiiiiiiiiiiiiiiiiiiiiiii

%======================================================================
\section{Feshbach molecule spectroscopy}\label{spectroscopy}
%======================================================================

The rich energy structure of Cs$_2$ Feshbach molecules requires
flexible methods for determining the molecular energy spectrum.
Previous studies on Feshbach molecules have mostly addressed the
last, most weakly bound state responsible for the respective
Feshbach resonance. Molecular binding energies have been measured
by applying various methods either to atomic
\cite{Donley2002amc,Claussen2003,Thompson2005,Ospelkaus2006uhm} or
to molecular samples
\cite{Regal2003cum,Bartenst2004b,Chin2005,Mark2007}.

%rapid magnetic field variations for an atomic sample near a
%Feshbach resonance \cite{Donley2002amc,Claussen2003}, by resonant
%excitation with sinusoidally oscillating magnetic fields
%\cite{Thompson2005}, by radio-frequency spectroscopy
%\cite{Ospelkaus2006uhm}

In this Section, we present our results on spectroscopy of weakly
bound trapped molecules. We use two different techniques to
measure the binding energies. Both techniques are suitable for
probing weakly as well as deeply bound molecular states.

The first method (Sec.~\ref{Sec_MM}) is based on a measurement of
the molecular magnetic moment \cite{Chin2005}. Magnetic moment
spectroscopy is a very general method, independent of selection
rules and wave function overlap requirements. It can be applied to
any molecular state and thus is an important tool for molecular
state identification. The method in particular allows us to follow
and investigate the avoided level crossings between different
molecular states. Transfers between different molecular states are
observed as sudden changes of the magnetic moment. In this way, we
are able to completely map out the molecular spectrum below the
atomic scattering continuum, including three $l$-wave states, two
of which had so far not been discovered.

%We note that
%the change in the magnetic moment was used in the first demonstrations of
%ultracold molecular clouds produced from atomic Bose-Einstein condensates
%\cite{Herbig2003,Duerr2004,Xu2003} to separate the molecules from the atoms.

The second method (Sec.~\ref{microwave}) uses microwave radiation
to measure binding energies of trapped molecules with very high
precision. We use a microwave pulse to drive a hyperfine
transition from a molecular bound state to a higher molecular
bound state that is associated with another channel of the
electronic ground-state manifold. Rapid spontaneous dissociation
loss \cite{Thompson2005b} provides the spectroscopic signal.

%----------------------------------------------------------------------
\subsection{Magnetic moment spectroscopy}
\label{Sec_MM}
%======================================================================

\begin{figure}
\includegraphics[width=8.5cm]{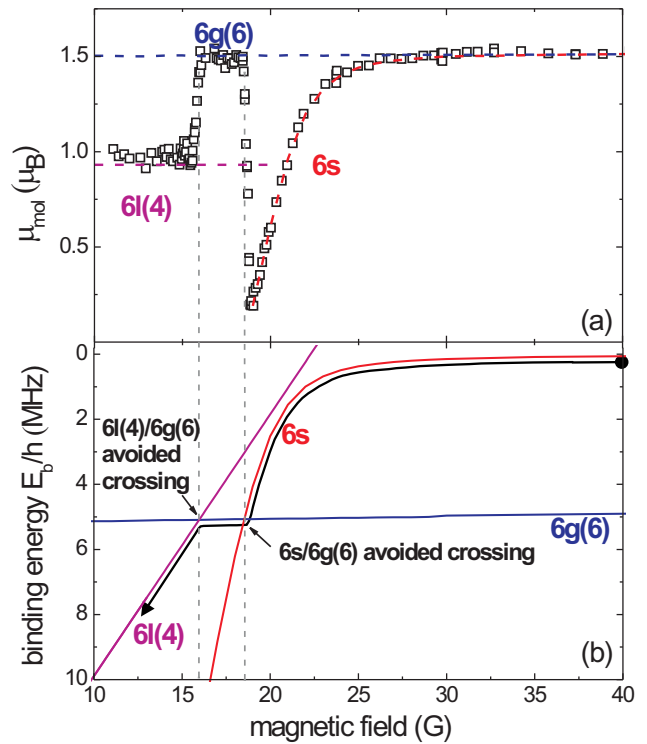}
\caption{(color online). Magnetic moment of Cs dimers across the
$6s-6g(6)-6l(4)$ molecular path. (a) The measured magnetic moments
(open squares) are compared to the NIST calculations (dashed
lines). The fast changes of the magnetic moment at 18.5\,G and
16\,G result from the $6l(4)/6g(6)$ and $6s/6g(6)$ avoided
crossings, respectively. (b) Molecular binding energies of the
$6s$, $6g(6)$ and $6l(4)$ levels calculated from the NIST model;
see also Fig.\,\ref{fig_roadmap}. The molecular path followed in
the measurement is indicated by the black arrow. \label{figmm}}
\end{figure}

%----------------------------------------------------------------------
\subsubsection{Bare energy levels}
%----------------------------------------------------------------------

\begin{table}
\caption{\label{tableMM} Measured magnetic moment $\mu_{\text{mol}}$
of Cs$_2$ molecules in different internal states with the
corresponding magnetic field range. The error of $\mu_{\text{mol}}$
accounts for the statistical error and a slight change of
$\mu_{\text{mol}}$ in the range considered. For each state also the
theoretical magnetic moment from the NIST model is listed.}
\begin{center}
\begin{ruledtabular}
\begin{tabular}{c|c c c c}
 Molecular  & $B$(G) & &$\mu_{\text{mol}}/\mu_B$&\\
state&& measured &&  NIST model \\
\hline
$6l(3)$ &4 - 9&0.75(4)&& 0.702  \\
        &12-16&0.75(2) && 0.702  \\
$4g(4)$ &5.5 - 12 &0.95(4) && 0.912 \\
        &15.5 - 20 &0.949(6) && 0.932  \\
$6l(4)$ &11 - 15&0.98(3) && 0.931 \\
        &16 - 24&0.96(1) && 0.931    \\
$6s$    &19& 0.192 && 0.191 \\
        & 45 & 1.519 && 1.515   \\
$6l(5)$ &15.5 - 23.5& 1.15(3) && 1.155 \\
        &26 - 37& 1.15(2) && 1.155   \\
$4d$    &41 - 43.2 &0.39(1) && 0.310 \\
        &45.5 - 47.1& 0.36(2) && 0.310 \\
$2g$    &51.7 - 52.2& 0.05(3) && 0.001  \\
$6g(6)$ &19 - 24 &1.49(1) && 1.503 \\
        &26 - 40& 1.5(1) && 1.503   \\
\end{tabular}
\end{ruledtabular}
\end{center}
\end{table}

We measure the molecular magnetic moment using the Stern-Gerlach
effect. Optically trapped molecules are initially prepared in a
single quantum state at a certain magnetic field $B$ by following
the procedure described in Sec.~\ref{preparation}.  The molecular
sample is then released from the trap. It starts to expand while
simultaneously a vertical magnetic field gradient $B' =
\partial B/ \partial z$ of typically 13\,G/cm is turned on. During
the time of flight, both the gravitational and the magnetic force
displace the center-of-mass position of the molecular cloud along
the vertical direction. The magnetic force acting on the molecules
is given by
\begin{equation}
F_z=\mu_{\text{mol}}B',
\end{equation}
where $\mu_{\text{mol}}$ is the molecular magnetic moment.  The
vertical relative displacement $\Delta z_{\text{mol}}$ of the
molecular cloud with respect to the position after expansion at
zero magnetic gradient is proportional to $\mu_{\text{mol}}$,
\begin{equation}
\Delta z_{\text{mol}}=\frac{1}{2}\frac{\mu_{\text{mol}}
B'}{m_{\text{mol}}} t_{\rm{SG}}^2,
\end{equation}
where $m_{\text{mol}}=2 m_{\text{at}}$ is the molecular mass and
$t_{\rm{SG}}$ is the time spent by the molecules in the magnetic
field gradient during the Stern-Gerlach procedure.

To minimize uncertainties resulting from $B'$, $t_{\rm{SG}}$, and
the spatial calibration of the imaging system, it is convenient to
measure $\mu_{\text{mol}}$ relative to the well-known magnetic
moment $\mu_\text{at}$ of the atoms. Consequently,
$\mu_{\text{mol}}$ can be written as
\begin{equation}
\mu_{\text{mol}}=\frac{\Delta z_{\text{mol}}}{\Delta
z_{\text{at}}}2\mu_\text{at},
\end{equation}
where $\Delta z_{\text{at}}$ is the measured displacement of atoms
for the same $B^{'}$ and $t_{\rm{SG}}$.

In previous experiments, we have determined $\mu_{\text{mol}}$ by
measuring the magnetic field gradient needed to levitate the
molecules against gravity \cite{Herbig2003,Chin2005}. For each
magnetic field value $B$, the value of $B'$ was adjusted to
maintain the levitation condition. This method is not practical
when $B$ is changed over a wide range. In the present experiments,
we measure the displacement of the molecular gas for a fixed $B'$
and for $B'=0$.

Once the magnetic moment $\mu_{\text{mol}}$ is known as a function
of $B$, the molecular binding energy $E_b$ is calculated by
integrating
\begin{equation}
\frac{\partial E_b}{\partial B} = 2\mu_{\text{at}}-\mu_{\text{mol}}.
\label{eqnmm}
\end{equation}
The integration constant is fixed by the atomic scattering
threshold where $E_b=0$. Eq.\,\ref{eqnmm} establishes a one-to-one
correspondence between $\mu_{\text{mol}}$ and  $E_b$ at each
magnetic field.

\begin{figure*}
\begin{center}
\includegraphics[scale=1]{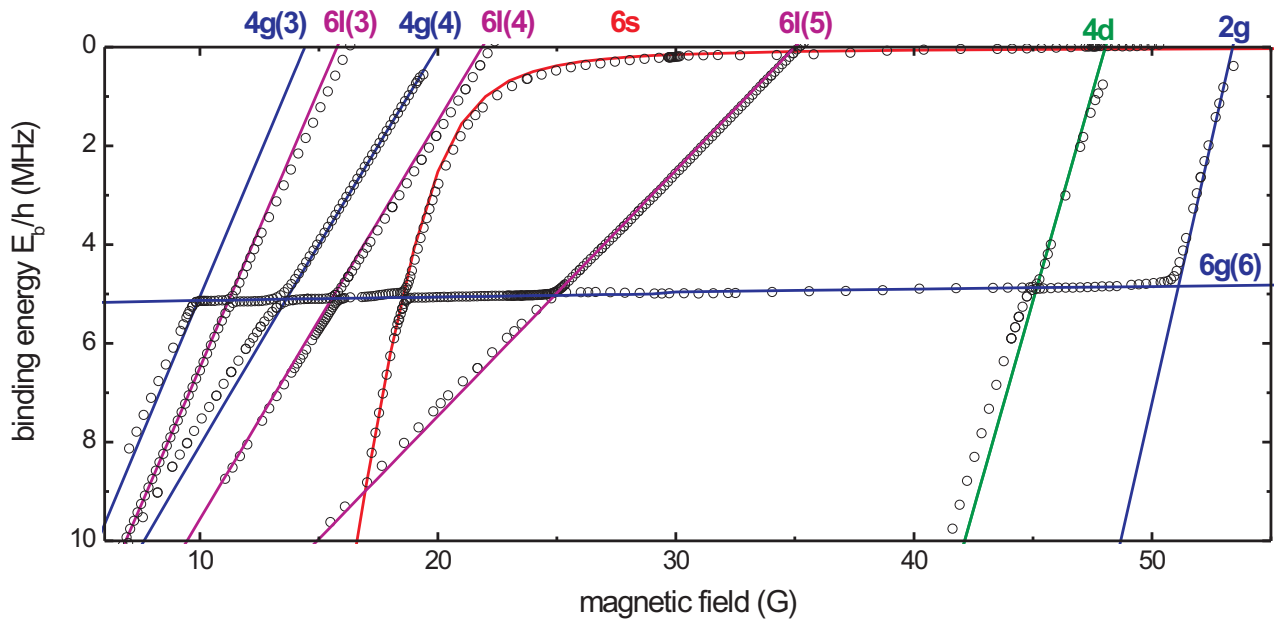}
\caption{(color online). Energy spectrum of weakly bound Cs
molecules as a function of the magnetic field. The binding
energies for the different molecular states are derived from
magnetic moment spectroscopy (open circles). The solid lines are
the molecular binding energies resulting from the extended NIST
model (for details see Sec.\,\ref{Sec_specialCs}). \label{figEb}}
\end{center}
\end{figure*}

An example of a magnetic moment measurement is shown in
Fig.\,\ref{figmm}. We produce $6s$ molecules from the 47.9 G
resonance, as discussed in Sec.~\ref{creation}. We then follow the
path indicated by the arrow in Fig.\,\ref{figmm}(b) and measure
the molecular magnetic moment at different values of $B$. We
observe a strong magnetic field dependence of the magnetic moment
of $6s$ molecules. Above 30\,G, where the $6s$ level runs almost
parallel to the atomic threshold (see Fig.\,\ref{figmm}(b)),
$\mu_{\text{mol}}$ is nearly constant with a value close to
2$\mu_{\text{at}}$ ($=\!1.5$\,$\mu_B$, with Bohr's magneton
$\mu_B$, for which $\mu_B/h \approx 1.400$\,MHz/G). When lowering
$B$ below 30\,G, we start to observe a decrease of
$\mu_{\text{mol}}$, which drops to one tenth of the initial value
within a magnetic field range of about 10\,G. This behavior is
explained by the strong coupling between two different $6s$
states. When further lowering the magnetic field,
$\mu_{\text{mol}}$ suddenly changes from 0.19\,$\mu_B$ to
1.5\,$\mu_B$ as the molecules are transfered to the $6g(6)$ state
via the $6s/6g(6)$ avoided crossing. The $6g(6)$ state has a
nearly constant magnetic moment, slightly less than
$\!1.5$\,$\mu_B$. Upon further lowering of $B$ the next avoided
crossing (to the state $4g(4)$, see Fig.\,\ref{fig_roadmap}) would
be expected at $13.6$\,G \cite{Chin2005,Mark2007}. However,
$\mu_{\text{mol}}$ undergoes a rapid change to a value of about
1\,$\mu_B$ at $\approx 16$\,G. This indicates the presence of a
new avoided crossing and hence the presence of a new state. The
existence of this state cannot be explained within the original
NIST model \cite{Leo2000,Chin2004}, which includes molecular
states only up to $g$-waves. The extension of the model to higher
order molecular states (Sec.~\ref{Sec_specialCs}) identifies this
state as a $6l(4)$ state \cite{TiesingaPrivate}.

Similar measurements have been performed for most of the molecular
states in the magnetic field range from 5 to 55\,G. The results of
our magnetic moment spectroscopy are summarized in
Table\,\ref{tableMM} and the molecular energy spectrum derived
using Eq.\,(\ref{eqnmm}) is shown in Fig.\,\ref{figEb} (open
circles) along with the results of the extended NIST model (solid
lines). We detect all the $s$, $d$-, $g$- and $l$-wave states in
the range of interest. Note that there are no $i$-wave states in
this range. All $d$-, $g$- and $l$-wave states exhibit a rather
constant magnetic moment. Consequently, we find a nearly linear
dependence of the binding energy on $B$, as shown in
Fig.\,\ref{figEb}.

In Table\,\ref{tableMM} and Fig.\,\ref{figEb} we compare our
results with the NIST model. In general, we find good agreement
with the theoretical predictions for the binding energies and
magnetic moments of the $s$, $d$ and $g$-wave states. The small
discrepancies observed for the lower branch of the $4g(4)$ state
and for the $4d$ state are probably the result of the more
complicated production schemes introducing larger systematic
errors in the measurements.

An important result of the magnetic moment spectroscopy is the
detection and characterization of three $l$-wave states, the
states $6l(3)$, $6l(4)$ and $6l(5)$. Recently, signatures of the
$6l(3)$ state have been reported in Ref.\,\cite{Mark2007}, whereas
the other two states had so far not been discovered. In contrast
to the $s$, $d$ and $g$-wave states, the $l$-wave states do not
reveal themselves via Feshbach resonances in atomic scattering as
the coupling to the atomic scattering state is too weak. Therefore
these states had previously not been included in the NIST model.
The extended NIST model shows the existence of these three
$l$-wave states and predicts their magnetic moments. Despite the
accuracy for these predictions, the model is not able to precisely
determine the binding energies. Our measurements now completely
characterize the three $l$-wave states and in particular give a
value for the binding energy at zero magnetic field where all
three states are degenerate. We find the binding energy of the
$6l$ manifold of states at zero magnetic field to be
17.61(9)\,MHz. In Fig.\,\ref{fig_roadmap} and in Fig.\,\ref{figEb}
we have down-shifted the NIST prediction of the $6l$ states by
$\approx 2.25$\,MHz to match the experimentally obtained binding
energies. The measurements also locate the magnetic field
positions where the three $6l$ states intersect the atomic
scattering continuum. We find the crossing positions for the
$6l(3)$, $6l(4)$, and $6l(5)$ states at 16.1(2) G, 22.0(2) G, and
35.0(2) G, respectively.

%----------------------------------------------------------------------
\subsubsection{Avoided crossings}\label{sec_avoidedcross}
%----------------------------------------------------------------------

\begin{figure}
\includegraphics[width=8.5cm]{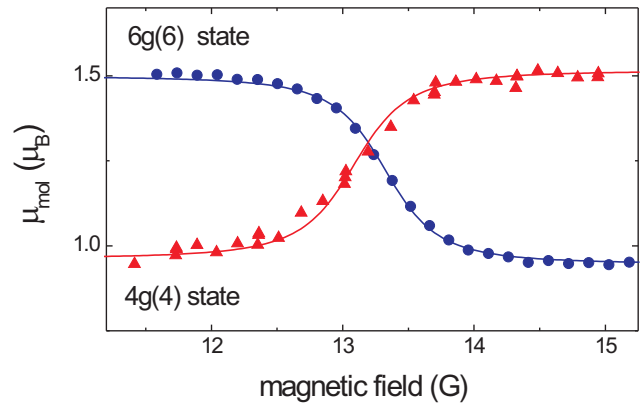}
\caption{(color online). Magnetic moments of Cs dimers across the
$6g(6)/4g(4)$ avoided crossing. Both the change in magnetic moment
from the $6g(6)$ to the $4g(4)$ state (circles) and the one from
the $4g(4)$ to the $6g(6)$ state (triangles) are shown. The
measured magnetic moments are fitted using Eq.\,(\ref{MMCoupling})
(solid lines). \label{figMMAC}}
\end{figure}

\begin{table}
\caption{\label{tableAC} Avoided crossing positions $B_0$ and
coupling strengths $V/h$ between the $6g(6)$ state and the
intersecting $f\ell(m_f)$ molecular states obtained by fitting the
measured magnetic moments with Eq.\,(\ref{MMCoupling}). The errors
are the one-sigma statistical uncertainties. $V/h$ measured with
different techniques are also reported (see notes).}
\begin{center}
\begin{ruledtabular}
\begin{tabular}{c|c|cc}
\vspace{3 pt} $f\ell(m_f)$  &  $B_0$ (G) & \multicolumn{2}{c}{$V/h$\,(kHz)} \\
\hline
$6l(3)$      &11.22(2)&  & 16(3)\footnotemark[1], 14(1)\footnotemark[2]  \\
$4g(4)$      &13.29(4)& 164 (30) & 150(10)\footnotemark[3] \\
$6l(4)$      &15.50(3)& 64(13)\footnotemark[4] & \\
$6l(5)$      &25.3(1)& 63(22)\footnotemark[4] & \\
$4d$         &45.15(4)&120(21) &   \\
\end{tabular}
\end{ruledtabular}\end{center}
\footnotetext[1]{Landau-Zener method.}
\footnotetext[2]{Interferometric method\,\cite{Mark2007}.}
\footnotetext[3]{Magnetic levitation method\,\cite{Chin2005}.}
\footnotetext[4]{The values should be considered as upper limits.}
\end{table}

Magnetic moment spectroscopy also allows a direct observation of
the avoided crossings between different molecular states. As is
well known, the coupling $V$ between two generic molecular states,
state 1 and state 2, modifies the bare energies $E_1$ and $E_2$ by
opening an energy gap $2V$ at the crossing position. In the limit
of a coupling strength $V$ that is small compared to the energy
separation to all other states, the avoided crossings can be
studied within a simple two-state model. This model takes the two
interacting bound states into account while both the couplings
with the scattering continuum and with other molecular states are
neglected. The coupled energy levels are given by
\begin{equation}
\label{solutionCoupling} E_\pm=\frac{(E_{1}+E_{2})\pm
\sqrt{(E_{1}-E_{2})^2+4V^2}}{2}.
\end{equation}
The energies $E_+$ and $E_-$ refer to the upper and lower adiabatic
levels of the avoided crossing. The derivatives $-\partial
E_\pm/\partial B$ correspond to the magnetic moments $\mu_+$ and
$\mu_-$ of the coupled states with
\begin{equation}
\label{MMCoupling} \mu_\pm=\frac{1}{2}(\mu_1+\mu_2)\mp \frac{1}{2}
\frac{(\mu_{2}-\mu_{1})^2 (B-B_0)}
{\sqrt{(\mu_{2}-\mu_{1})^2(B-B_0)^2+4V^2}}.
\end{equation}
Here, $B_0$ is the magnetic field at the avoided-crossing
position, and $\mu_1$ and $\mu_2$ are the magnetic moments of the
two bare molecular states.

In the following we focus on the avoided crossings between the
$6g(6)$ state and the other $f\ell(m_f)$ states.
Fig.\,\ref{figMMAC} shows the magnetic moments $\mu_+$ (circles)
and $\mu_-$ (triangles) across the $6g(6)/4g(4)$ avoided crossing.
To derive the coupling strength between these two states, we fit
our data using Eq.\,(\ref{MMCoupling}) by leaving $\mu_1$,
$\mu_2$, $B_0$, and $V$ as free parameters. The same procedure is
adopted to analyze the other crossings. The coupling strengths and
the avoided crossing positions are listed in Table\,\ref{tableAC}.
For comparison, we include in Table\,\ref{tableAC} measurements of
$V$ obtained with other techniques, such as the Landau-Zener
method discussed below, a magnetic levitation
method\,\cite{Chin2005}, and an interferometric
method\,\cite{Mark2007}.

\begin{figure}
\includegraphics[width=8.5cm]{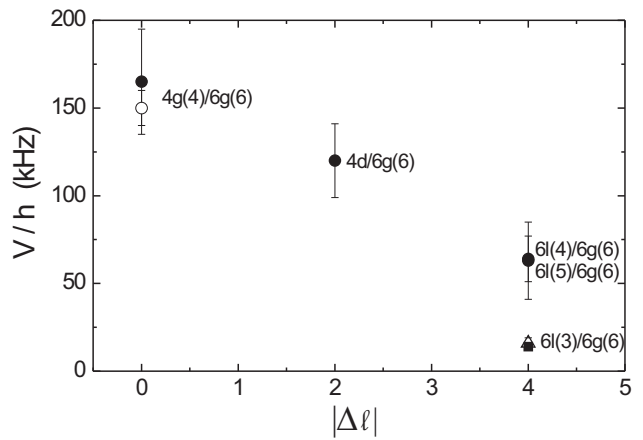}
\caption{Coupling strengths $V/h$ between the $6g(6)$ state and the
intersecting $f\ell(m_f)$ molecular states as a function of the
difference their orbital angular momentum quantum numbers, $|\Delta
\ell|$. The data refer to the values obtained via the magnetic
moment spectroscopy (filled circles), a magnetic levitation method
(empty circle)\,\cite{Chin2005}, an interferometer method
(square)\,\cite{Mark2007}, and the Landau-Zener method (triangle).
\label{figdeltal}}
\end{figure}

In Fig.\,\ref{figdeltal} we plot the measured coupling strengths
$V$ between the $6g(6)$ state and the other intersecting states as
a function of the difference in orbital angular momentum $|\Delta
\ell|$. While the $6g(6)/4g(4)$ and the $6g(6)/4d$ crossings are
the result of the first order spin-spin dipole interaction, the
crossings with the l-wave states are second order. As a general
trend, crossings with larger $|\Delta \ell|$ tend to have a weaker
coupling.

%We observe a decrease of $V$ for an increasing $|\Delta \ell|$.
%This general trend can be explained by the different coupling
%mechanisms involved. In particular, states with different $\ell$
%couple via the weak spin-spin dipole and second-order spin-orbit
%interaction while the coupling between states with $\Delta \ell=0$
%is essentially mediated by the exchange and van der Waals
%interactions \cite{Koehler2006}.

\begin{figure}
\includegraphics[width=8.5cm]{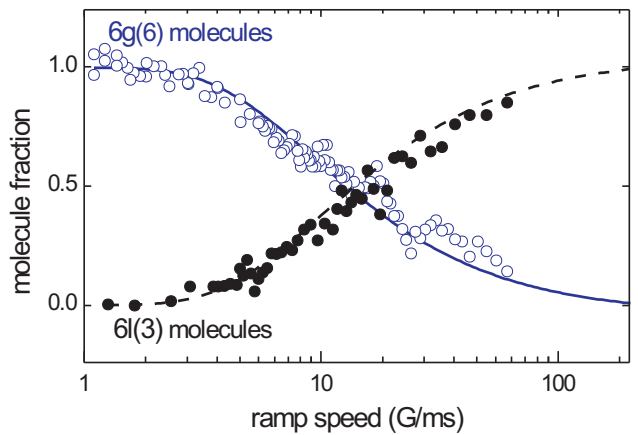}
\caption{(color online). Conversion efficiency on the
$6g(6)/6l(3)$ avoided crossing as a function of the ramp speed. A
pure sample of $6l(3)$ molecules is partially transferred into the
$6g(6)$ state at different ramp speeds. We measure either the
fraction of transferred $6g(6)$ molecules (open circles) or the
fraction of non-converted $6l(3)$ molecules (filled circles). The
solid line refers to the Landau-Zener formula for $p$ given by
Eq.\,(\ref{EqAc}), while the dashed line is $1-p$. \label{figAC}}
\end{figure}

Systematic errors in our avoided crossing measurements stem from
the finite size of the molecular cloud and the change of
$\mu_{\text{mol}}$ during the free fall and expansion. These
effects cause an apparent broadening of the avoided crossings and
lead to an overestimation of the coupling strengths, in particular
for the narrower crossings.  We find a limit on the minimum
coupling strength that can be extracted with reasonable precision.
We estimate from simulations that coupling strengths below
$h\times 50$\,kHz can no longer be sensitively measured with our
present method.

An alternative method to determine the coupling strengths of
avoided crossings is based on the Landau-Zener tunneling model
\cite{Landau1932xxx,Zener1932xxx,Julienne2004}, already discussed
in Sec.\,\ref{state_transfer}. Eq.\,(\ref{eq_Bcrit}) shows a
quadratic dependence of the critical ramp speed $r_c$ on the
coupling strength $V$. The probability to transfer molecules from
one bare state to the next in a single passage through the avoided
crossing is given by the well-known Landau-Zener formula
\cite{Landau1932xxx,Zener1932xxx}
\begin{equation}
\label{EqAc} p=1-\exp{\left(-r_c/\dot{B}\right)},
\end{equation}
where $\dot{B}$ is the ramp speed.

As an example, we apply this method to the 6$g$(6)/6$l$(3) avoided
crossing. We measure the conversion efficiency of molecules from
the $6l(3)$ state below the crossing to the $6g(6)$ state by
sweeping the magnetic field across the $6g(6)/6l(3)$ crossing at
various ramp speeds $\dot{B}$. The results are shown in
Fig.\,\ref{figAC}. For $\dot{B}\ll r_{c}$, the molecules are
adiabatically transferred to the $6g(6)$ state (open circles)
whereas, for $\dot{B}\gg r_{c}$, they end up in the $6l(3)$ state
above the crossing. The conversion efficiency is measured by
detecting the $6g(6)$ molecules (open circles) and also by
detecting the $6l(3)$ molecules (filled circles). By fitting our
data with Eq.\,(\ref{EqAc}), we estimate the coupling strength of
the $6l(3)/6g(6)$ crossing to be $V=h\times16(3)$\,kHz. This value
is consistent with the result of 14(1)\,kHz obtained in
Ref.\,\cite{Mark2007} using a more precise interferometric
technique.

%----------------------------------------------------------------------
\subsection{Microwave spectroscopy}
\label{microwave}
%======================================================================

\begin{figure}
\includegraphics[width=8.5cm]{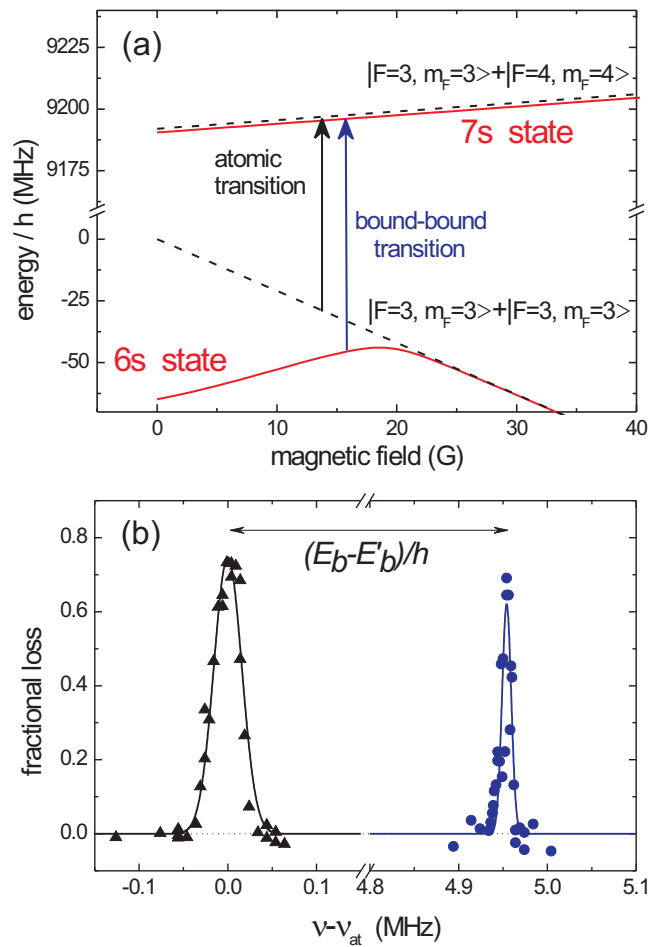}
\caption{(color online). Microwave spectroscopy of Cs dimers. (a)
A bound-bound transition is driven from the $6s$ state to a $7s$
state slightly below the $\vert F\!=\!3, m_{F}\!=\!3\rangle +
\vert F\!=\!4, m_{F}\!=\!4\rangle$ scattering channel, as
illustrated by the longer arrow. The $7s$ state is offset from the
scattering channel for clarity. The frequency corresponding to the
$\vert F\!=\!3, m_{F}\!=\!3\rangle\!\rightarrow\!\vert F\!=\!4,
m_{F}\!=\!4\rangle$ atomic transition at zero-field is
$\nu_{\text{at}}\approx 9.193$\,GHz. (b) Microwave spectrum of
atoms (triangles) and $6s$ molecules (dots) at $B\approx 18.7\,G$
as a function of the frequency offset $\nu-\nu_{\text{at}}$. The
molecular transition corresponds to a sharp loss resonance. We
determine the center position to be 4.9545(3)\,MHz and the
resonance width to 12(3)\,kHz from a gaussian fit (solid line).
\label{MW}}
\end{figure}

Molecules in the $6s$ state (see Fig.\,\ref{fig_roadmap}) are of
particular interest as {\em quantum halo} states
\cite{Jensen2004}. Halo states are extremely weakly bound dimers
characterized by a large interatomic separation that greatly
exceeds the van der Waals length  $r_0$ (for Cs, $r_0\simeq 101
\, a_0$) and by a binding energy much smaller than the van der
Waals energy (for Cs, $\hbar^2/m r_0^2\approx h \times
2.708$\,MHz) \cite{Koehler2006}. These states are universal in the
sense that they are fully characterized by a large atomic $s$-wave
scattering length $a$. In particular, the wave function does not
depend on the microscopic details of the scattering potential.
%All these features make the halo states crucial in
%understanding the two-body universal physics and in Efimov-type states.
The precise knowledge of the $6s$ state is crucial for
understanding universal two-body physics and for studying
universal three-body Efimov-type states \cite{Kraemer2006efe}.

We detect molecular transitions induced by microwave radiation to
probe the binding energy of the $6s$ molecules. The relevant
atomic states are the lowest hyperfine state $\vert F\!=\!3,
m_{F}\!=\!3\rangle$ and the doubly-polarized state $\vert F\!=\!4,
m_{F}\!=\!4\rangle$. Fig.\,\ref{MW}(a) shows the energy level
structure of the two scattering channels $\vert F\!=\!3,
m_{F}\!=\!3\rangle + \vert F\!=\!3, m_{F}\!=\!3\rangle$ and $\vert
F\!=\!3, m_{F}\!=\!3\rangle + \vert F\!=\!4, m_{F}\!=\!4\rangle$.
The bound states involved in the molecular transition are the $6s$
state and a $7s$ state that lies slightly below the atomic
scattering channel $\vert F\!=\!3, m_{F}\!=\!3\rangle + \vert
F\!=\!4, m_{F}\!=\!4\rangle$.

The weakly bound $7s$ state is directly related to the large
triplet scattering length $a_T$ that dominates the $\vert F\!=\!3,
m_{F}\!=\!3\rangle + \vert F\!=\!4, m_{F}\!=\!4\rangle$ scattering
channel. The Cs triplet scattering length, predicted by the NIST
model, is $(2400 \pm 100) a_0$, and consequently the $7s$ state
has a small binding energy of $E'_{b}=\hbar^2/m a_T^2\approx h
\times 5$\,kHz.

We map out the binding energy of the $6s$ molecules by measuring
the transition frequency $\nu_{\text{mol}}$ from the $6s$ to the
$7s$ state as a function of $B$. The binding energy is then given
by
\begin{equation}
\label{mw_freq} E_b(B)=h \times
(\nu_{\text{mol}}(B)-\nu_{\text{at}}(B))+E'_b,
\end{equation}
where $\nu_{\text{at}}(B)$ is the $\vert F\!=\!3,
m_{F}\!=\!3\rangle\!\rightarrow\!\vert F\!=\!4,
m_{F}\!=\!4\rangle$ atomic transition, which follows the
Breit-Rabi formula and is used here as frequency reference. In our
experiment, we again start with optically trapped $6s$ molecules
at some magnetic field $B$. A microwave pulse of typically 5\,ms
duration drives the bound-bound transition, and partially
transfers molecules from the $6s$ state to the $7s$ state. We then
hold the sample in the trap for 10\,ms and we detect the total
number of remaining molecules using the techniques described in
Sec.\,\ref{detection}. We perform similar measurements at
different magnetic fields to recover $E_b$ within the magnetic
field range of investigation.

As a frequency reference, we measure $\nu_{\text{at}}(B)$ on a
trapped sample of $4 \times 10^5$ Cs atoms at $T \approx 200$\,nK,
initially prepared in the hyperfine ground state $\vert F\!=\!3,
m_{F}\!=\!3\rangle$. For each $B$, we apply a microwave pulse
resonant to the $\vert F\!=\!3, m_{F}\!=\!3\rangle\rightarrow
\vert F\!=\!4, m_{F}\!=\!4\rangle$ hyperfine transition. The atoms
are then detected after a holding time in the trap of typically
100\,ms. The microwave excitation results in resonant loss from
the atomic sample.

The particle losses observed in both the atomic and the molecular
sample are the result of hyperfine spin relaxation
\cite{Thompson2005b,Koehler2005}. In the atomic case, the
relaxation is driven by the binary collision of two free atoms,
while in the molecular case it can be considered as being driven
by a collision within the molecule \cite{Koehler2005}, leading to
spontaneous dissociation. In any case, one of the atoms is subject
to a spin flip, releasing the hyperfine energy that greatly
exceeds the trap depth and leading to trap loss. According to the
NIST model, the $7s$ state is coupled to several possible decay
channels, causing a decay width of the state of
$\sim2\pi\times$70\,Hz \cite{TiesingaPrivate}. We in fact observe
a decay of $7s$ molecules on a timescale of a few ms. In the case
of Cs atoms in the $\vert F\!=\!3, m_{F}\!=\!3\rangle + \vert
F\!=\!4, m_{F}\!=\!4\rangle$ scattering channel, we measure a
lifetime of $\sim$ 50\,ms consistent with the predicted two-body
loss coefficient of $5\times
10^{-12}$\,cm$^3$/s\,\cite{TiesingaPrivate}.

%Similarly to the case of $^{85}$Rb$_2$
%molecules\,\cite{Thompson2005b}, this high loss rate owes to
%spontaneous dissociation of the $7s$ molecules, which does not
%preserve the total $m_f$. This mechanism \cite{Koehler2005} refers
%to the situation in which one of the components in the molecule
%experiences a spin flip, which causes the molecules to dissociate.

\begin{figure}
\includegraphics[width=8.5cm]{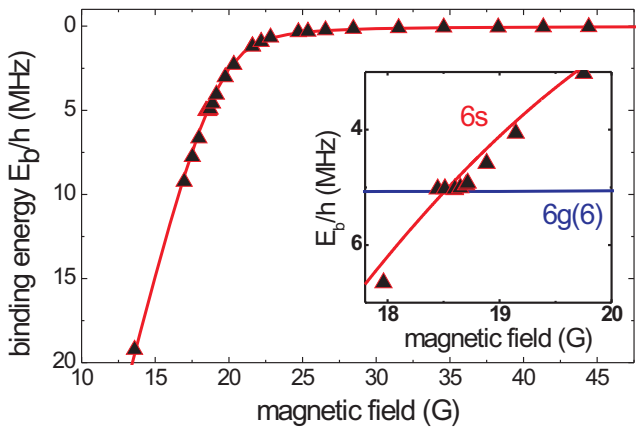}
\caption{(color online). Binding energy of the $6s$ molecules as a
function of the magnetic field (triangles). The binding energies
correspond to the measured frequency shift from the expected
$\vert F\!=\!3, m_{F}\!=\!3\rangle\rightarrow \vert F\!=\!4,
m_{F}\!=\!4\rangle$ atomic transition (see Eq. (\ref{mw_freq})).
The statistical errors are about 1 kHz, i.e. much smaller than the
data symbol size. The solid line is the result of the NIST model.
The inset is an expanded view of the binding energy in the
proximity of the $6s/6g(6)$ avoided crossing. \label{MWtot}}
\end{figure}

A typical microwave spectrum for both atoms and molecules is shown
in Fig.\,\ref{MW}(b). The resonant frequencies and the line widths
are determined by fitting the data with gaussian profiles. The
molecular transition shows a narrow and symmetric loss resonance.
From the fit, we find a line width of 12(3)\,kHz. This value is
close to our experimental resolution of $\sim 10$\,kHz,
essentially resulting from magnetic field fluctuations. As
demonstrated in Refs.\,\cite{Bartenst2004b,Chin2005PRA}, the
symmetry of line shape indicates that a bound-bound transition
occurs, even in the presence of magnetic field broadening. We
cannot distinguish bound-bound from possible bound-free
transitions due to the small energy difference between the $\vert
F\!=\!3, m_{F}\!=\!3\rangle + \vert F\!=\!4, m_{F}\!=\!4\rangle$
scattering channel and the $7s$ state. However, we believe that
the bound-bound transition dominates as the transition probability
for a bound-free transition is expected to be much weaker due to
the smaller Franck-Condon overlap between the initial and final
state\,\cite{Chin2005PRA}.

Figure\,\ref{MWtot} shows the binding energies of $6s$ molecules
in a magnetic field range from 12 to 45 G together with the
predictions from the NIST model. The inset shows an expanded view
of the binding energy in the proximity of the $6s/6g(6)$ avoided
crossing at around $18.5$\,G.  We observe an increase of the
microwave power needed to drive the  bound-bound transition when
the avoided crossing is approached. The $6s$ and $6g(6)$ state
couple and the molecules are in a dressed state. A microwave pulse
can drive molecular transitions that change the total angular
momentum $f$ and its projection $m_f$, while the orbital momentum
$\ell$ has to be conserved. The bound-bound transition between the
$6g(6)$ and the $7s$ state with $\Delta \ell=4$ is hence
forbidden.

The microwave measurements on the $6s$ state provide precise
binding energies of up to about $h\times$\,20\,MHz. Higher binding
energies can in principle be accessed by further lowering the
magnetic field. The comparison between our results and the NIST
model generally shows very good agreement. We have observed small
deviations between theory and experiment when the $6s$ state
starts to bend towards larger binding energies (see inset of
Fig.\,\ref{MWtot}). This deviation suggests that the $6s$ state is
perturbed by the coupling to other molecular states. Our data
provide high precision input for further refinements of the NIST
model.

%================================================================================
\section{Conclusion}
%================================================================================

We have explored the rich internal structure of weakly bound
Cs$_2$ Feshbach molecules, prepared in a CO$_2$-laser trap.
Magnetically induced association based on three different Feshbach
resonances served as the entrance door into the manifold of
molecular states. We have developed a set of methods to transfer
molecules to various internal states, to clean the population in
the optical trap from remaining atoms and from molecules in
unwanted states, and to detect the molecular population via
controlled dissociation. In particular, we have investigated so
far unexplored $l$-wave states, for which direct Feshbach
association is not possible because of negligible coupling to
atomic scattering states.

We have determined the binding energy spectrum using two different
techniques. %both a technique based on their different magnetic
%moments and on hyperfine molecular transitions.
Magnetic moment spectroscopy has been demonstrated as a versatile
and sensitive method to detect molecular states. It shows avoided
crossings between different molecular states and reveals the
presence of higher partial wave states. Using this technique we
have mapped out the molecular spectrum up to binding energies of
$E_b/h = 10$ MHz and in a magnetic field range from 5 to 55\,G.
Using microwave spectroscopy, we have performed highly precise
measurements of the binding energy of a particularly important
$s$-wave state above 13\,G, where $E_b/h < 20$\,MHz. The results
show how this state, which essentially determines the $s$-wave
scattering length, evolves into a weakly bound state with
quantum-halo character. These results are important for
applications of this $s$-wave state to universal few-body quantum
physics, such as the exploration of Efimov states.

Our measurements provide a sensitive test for the theoretical NIST
model, which was developed to describe quantum scattering
phenomena of Cs atoms. We could confirm the basic predictions of
this model on the weakly bound molecular structure. The
exploration of novel $l$-wave states and highly precise
measurements on a weakly bound $s$-wave state provide experimental
input for further refinements of the NIST model.

In a broader perspective, our work demonstrates general ways to
manipulate Feshbach molecules through elaborate magnetic-field
control. This extends the experimental tool-box available for the
preparation of homo- and heteronuclear ultracold molecules in
desired internal states.

\begin{acknowledgments}
We thank E.\ Tiesinga, P.\ Julienne, and C.\ Williams for
providing us with invaluable theoretical input and J. Hutson and
A. Simoni for helpful discussions. We acknowledge support by the
Austrian Science Fund (FWF) within SFB 15 (project part 16) and by
the European Union within the Cold Molecules TMR Network under
contract No.\,HPRN-CT-2002-00290. M.\,M.\ acknowledges support
within the Ph.D.\ program DOC of the Austrian Academy of Sciences,
and F.\,F. and C.\,C.\ within the Lise Meitner program of the FWF.
S.\,K.\ is supported by the European Community with a Marie Curie
Intra-European Fellowship.
\end{acknowledgments}

%================================================================================
\begin{appendix}* %---*-------------APPENDIX-------------------------%
%================================================================================

\section{magnetic field setup and calibration}\label{appendix}

The experiments with ultracold Cs molecules are performed in a
twelve-sided polygonal stainless-steel vacuum chamber with large
re-entrant viewports on top and bottom for maximum optical access
along the vertical axis \cite{Weber2003PhD}. The CO$_2$-laser
light is brought in along the horizontal plane through special
ZnSe-viewports. On the twelve sides of the polgygon, there are in
total 4 pairwise opposite viewports for the CO$_2$-laser light and
6 pairwise opposite viewports for the near-infrared laser cooling
and trapping light and for imaging. The two remaining opposite
openings are reserved for the atomic beam, the Zeeman slowing
laser beam, and the vacuum pumps. The total magnetic field is
produced by several sets of coils, some mounted on the steel
chamber, others placed inside the re-entrant viewports closer to
the trap center, but still outside the vacuum. The presence of the
metal limits the magnetic field switching times as a result of
eddy currents. Nevertheless we achieve a maximum of experimental
flexibility by combining the larger fields of the bigger coils
with the more rapidly switchable fields of the smaller coils
inside the viewports.

\subsection{Bias field} The vertical offset field for molecule production
and manipulation is created by a pair of water cooled coils with a
mean radius of $66$\,mm in approximate Helmholtz configuration.
The coils are placed inside the re-entrant viewports along the rim
of the windows. They allow a magnetic field of up to $60$\,G for
dc-operation with a typical $1/e$-switching time of $1.5$\,ms. The
current from a programmable power supply is servo-loop controlled.
A second pair of large coils with a mean radius of $112.5$\,mm is
attached to the outside of the flanges that hold the re-entrant
viewports and can provide an additional dc-field of up to
$200$\,G.

For the molecule transfer schemes as described in
Sec.\,\ref{preparation}, one further set of air cooled coils and a
single ``booster-coil'' are used inside the re-entrant viewport: A
pair of coils in approximate Helmholtz configuration with a radius
of $44$\,mm is mounted on plastic holders near the vaccum window
as close to the trap center as possible. The coils with a
servo-loop controlled current produce a magnetic field of up to
$10$\,G, while the $1/e$ switching time is $\sim 300$\,$\mu$s. As
a result, ramp speeds in the range of $30$\,G/ms can be achieved.
The fastest magnetic field changes are realized by a small, single
``booster'' coil with only 4 windings and a diameter of $24$\,mm
at a distance of $\sim30$\,mm to the trap center placed inside the
top re-entrant viewport. Using a capacitor bench and servo-loop
control for the current we achieve magnetic field pulses with
amplitudes of up to $7$\,G. The maximum pulse duration of $1$\,ms
is sufficiently long to adjust the offset field of the other coils
within this time. With a typical switching time of $400$\,ns we
achieve ramp speeds of up to 17000\,G/ms. To change the ramp speed
we vary the pulse amplitude as the rise-time cannot be adjusted.
The coupling of ramp speed and pulse amplitude is somewhat
problematic. It limits the possible ramp speeds at certain avoided
crossings, because too large pulse amplitudes can produce
uncontrolled ramps over other avoided crossings nearby. Note that
the booster coil also produces a magnetic field gradient. However,
his gradient is irrelevant for the experiments reported here.

The relative stability of the servo-loop controllers is about
$10^{-5}$ and thus well below the ambient magnetic line noise
($\sim 10$\,mG).

%For compensating the magnetic field of the earth and residual stray
%fields, three additional coil-pairs are setup in the experiment. For
%each direction one pair can produce a field of about $2$\,G.

\subsection{Gradient field} The magnetic gradient field is
produced by a pair of water cooled coils in approximate
anti-Helmholtz configuration. These coils with a radius of
$66$\,mm are also placed inside the re-entrant viewports. They
allow a dc-gradient field of up to $80$\,G/cm. Large field
gradients can be switched within $\sim 3$\,ms, limited by eddy
currents. For small gradients such as $13$\,G/cm, as used in the
magnetic moment spectroscopy measurements, we measure somewhat
lower magnetic switching times of $\sim1$\,ms.

%As described in Ref.\,\cite{Herbig2003}, when superimposing a bias
%field with the gradient field, the atoms (or molecules) experience a
%small magnetic force outward directed from the trap center.

\subsection{Magnetic field calibration} To calibrate the magnetic
field we use the microwave technique on a trapped atomic sample as
described in Sec.\,\ref{microwave}. We use the Breit-Rabi formula
to determine the magnetic field value from a measurement of the
atomic hyperfine transition frequency between the states $\vert
F\!=\!3, m_{F}\!=\!3\rangle \rightarrow \vert
F\!=\!4,m_F\!=\!4\rangle$. Line noise limits the stability of the
magnetic field to about $10$\,mG for typical integration times.

\end{appendix}

\bibliography{Spectroscopy_Mols_arXiv_Submission}

\begin{thebibliography}{53}
\expandafter\ifx\csname natexlab\endcsname\relax\def\natexlab#1{#1}\fi
\expandafter\ifx\csname bibnamefont\endcsname\relax
  \def\bibnamefont#1{#1}\fi
\expandafter\ifx\csname bibfnamefont\endcsname\relax
  \def\bibfnamefont#1{#1}\fi
\expandafter\ifx\csname citenamefont\endcsname\relax
  \def\citenamefont#1{#1}\fi
\expandafter\ifx\csname url\endcsname\relax
  \def\url#1{\texttt{#1}}\fi
\expandafter\ifx\csname urlprefix\endcsname\relax\def\urlprefix{URL }\fi
\providecommand{\bibinfo}[2]{#2}
\providecommand{\eprint}[2][]{\url{#2}}

\bibitem[{\citenamefont{K\"ohler et~al.}(2006)\citenamefont{K\"ohler,
  G\'{o}ral, and Julienne}}]{Koehler2006}
\bibinfo{author}{\bibfnamefont{T.}~\bibnamefont{K\"ohler}},
  \bibinfo{author}{\bibfnamefont{K.}~\bibnamefont{G\'{o}ral}},
  \bibnamefont{and} \bibinfo{author}{\bibfnamefont{P.~S.}
  \bibnamefont{Julienne}}, \bibinfo{journal}{Rev.\, Mod.\, Phys.\,}
  \textbf{\bibinfo{volume}{78}}, \bibinfo{pages}{1311} (\bibinfo{year}{2006}).

\bibitem[{\citenamefont{Donley et~al.}(2002)\citenamefont{Donley, Clausen,
  Thompson, and Wieman}}]{Donley2002amc}
\bibinfo{author}{\bibfnamefont{E.~A.} \bibnamefont{Donley}},
  \bibinfo{author}{\bibfnamefont{N.~R.} \bibnamefont{Clausen}},
  \bibinfo{author}{\bibfnamefont{S.~T.} \bibnamefont{Thompson}},
  \bibnamefont{and} \bibinfo{author}{\bibfnamefont{C.~E.}
  \bibnamefont{Wieman}}, \bibinfo{journal}{Nature}
  \textbf{\bibinfo{volume}{417}}, \bibinfo{pages}{529} (\bibinfo{year}{2002}).

\bibitem[{\citenamefont{Herbig et~al.}(2003)\citenamefont{Herbig, Kraemer,
  Mark, Weber, Chin, N\"{a}gerl, and Grimm}}]{Herbig2003}
\bibinfo{author}{\bibfnamefont{J.}~\bibnamefont{Herbig}},
  \bibinfo{author}{\bibfnamefont{T.}~\bibnamefont{Kraemer}},
  \bibinfo{author}{\bibfnamefont{M.}~\bibnamefont{Mark}},
  \bibinfo{author}{\bibfnamefont{T.}~\bibnamefont{Weber}},
  \bibinfo{author}{\bibfnamefont{C.}~\bibnamefont{Chin}},
  \bibinfo{author}{\bibfnamefont{H.-C.} \bibnamefont{N\"{a}gerl}},
  \bibnamefont{and} \bibinfo{author}{\bibfnamefont{R.}~\bibnamefont{Grimm}},
  \bibinfo{journal}{Science} \textbf{\bibinfo{volume}{301}},
  \bibinfo{pages}{1510} (\bibinfo{year}{2003}).

\bibitem[{\citenamefont{D\"{u}rr
  et~al.}(2004{\natexlab{a}})\citenamefont{D\"{u}rr, Volz, Marte, and
  Rempe}}]{Duerr2004mols}
\bibinfo{author}{\bibfnamefont{S.}~\bibnamefont{D\"{u}rr}},
  \bibinfo{author}{\bibfnamefont{T.}~\bibnamefont{Volz}},
  \bibinfo{author}{\bibfnamefont{A.}~\bibnamefont{Marte}}, \bibnamefont{and}
  \bibinfo{author}{\bibfnamefont{G.}~\bibnamefont{Rempe}},
  \bibinfo{journal}{Phys.\ Rev.\ Lett.} \textbf{\bibinfo{volume}{92}},
  \bibinfo{pages}{020406} (\bibinfo{year}{2004}{\natexlab{a}}).

\bibitem[{\citenamefont{Xu et~al.}(2003)\citenamefont{Xu, Mukaiyama,
  Abo-Shaeer, Chin, Miller, and Ketterle}}]{Xu2003}
\bibinfo{author}{\bibfnamefont{K.}~\bibnamefont{Xu}},
  \bibinfo{author}{\bibfnamefont{T.}~\bibnamefont{Mukaiyama}},
  \bibinfo{author}{\bibfnamefont{J.~R.} \bibnamefont{Abo-Shaeer}},
  \bibinfo{author}{\bibfnamefont{J.~K.} \bibnamefont{Chin}},
  \bibinfo{author}{\bibfnamefont{D.~E.} \bibnamefont{Miller}},
  \bibnamefont{and} \bibinfo{author}{\bibfnamefont{W.}~\bibnamefont{Ketterle}},
  \bibinfo{journal}{Phys.\ Rev.\ Lett.} \textbf{\bibinfo{volume}{91}},
  \bibinfo{pages}{210402} (\bibinfo{year}{2003}).

\bibitem[{\citenamefont{Regal et~al.}(2003)\citenamefont{Regal, Ticknor, Bohn,
  and Jin}}]{Regal2003cum}
\bibinfo{author}{\bibfnamefont{C.~A.} \bibnamefont{Regal}},
  \bibinfo{author}{\bibfnamefont{C.}~\bibnamefont{Ticknor}},
  \bibinfo{author}{\bibfnamefont{J.~L.} \bibnamefont{Bohn}}, \bibnamefont{and}
  \bibinfo{author}{\bibfnamefont{D.~S.} \bibnamefont{Jin}},
  \bibinfo{journal}{Nature} \textbf{\bibinfo{volume}{424}}, \bibinfo{pages}{47}
  (\bibinfo{year}{2003}).

\bibitem[{\citenamefont{Strecker et~al.}(2003)\citenamefont{Strecker,
  Partridge, and Hulet}}]{Strecker2003coa}
\bibinfo{author}{\bibfnamefont{K.~E.} \bibnamefont{Strecker}},
  \bibinfo{author}{\bibfnamefont{G.~B.} \bibnamefont{Partridge}},
  \bibnamefont{and} \bibinfo{author}{\bibfnamefont{R.~G.} \bibnamefont{Hulet}},
  \bibinfo{journal}{Phys. Rev. Lett.} \textbf{\bibinfo{volume}{91}},
  \bibinfo{eid}{080406} (\bibinfo{year}{2003}).

\bibitem[{\citenamefont{Cubizolles et~al.}(2003)\citenamefont{Cubizolles,
  Bourdel, Kokkelmans, Shlyapnikov, and Salomon}}]{Cubizolles2003pol}
\bibinfo{author}{\bibfnamefont{J.}~\bibnamefont{Cubizolles}},
  \bibinfo{author}{\bibfnamefont{T.}~\bibnamefont{Bourdel}},
  \bibinfo{author}{\bibfnamefont{S.~J. J. M.~F.} \bibnamefont{Kokkelmans}},
  \bibinfo{author}{\bibfnamefont{G.~V.} \bibnamefont{Shlyapnikov}},
  \bibnamefont{and} \bibinfo{author}{\bibfnamefont{C.}~\bibnamefont{Salomon}},
  \bibinfo{journal}{Phys. Rev. Lett.} \textbf{\bibinfo{volume}{91}},
  \bibinfo{pages}{240401} (\bibinfo{year}{2003}).

\bibitem[{\citenamefont{Jochim et~al.}(2003{\natexlab{a}})\citenamefont{Jochim,
  Bartenstein, Altmeyer, Hendl, Chin, {Hecker Denschlag}, and
  Grimm}}]{Jochim2003pgo}
\bibinfo{author}{\bibfnamefont{S.}~\bibnamefont{Jochim}},
  \bibinfo{author}{\bibfnamefont{M.}~\bibnamefont{Bartenstein}},
  \bibinfo{author}{\bibfnamefont{A.}~\bibnamefont{Altmeyer}},
  \bibinfo{author}{\bibfnamefont{G.}~\bibnamefont{Hendl}},
  \bibinfo{author}{\bibfnamefont{C.}~\bibnamefont{Chin}},
  \bibinfo{author}{\bibfnamefont{J.}~\bibnamefont{{Hecker Denschlag}}},
  \bibnamefont{and} \bibinfo{author}{\bibfnamefont{R.}~\bibnamefont{Grimm}},
  \bibinfo{journal}{Phys. Rev. Lett} \textbf{\bibinfo{volume}{91}},
  \bibinfo{pages}{240402} (\bibinfo{year}{2003}{\natexlab{a}}).

\bibitem[{\citenamefont{Jochim et~al.}(2003{\natexlab{b}})\citenamefont{Jochim,
  Bartenstein, Altmeyer, Hendl, Riedl, Chin, Denschlag, and
  Grimm}}]{Jochim2003}
\bibinfo{author}{\bibfnamefont{S.}~\bibnamefont{Jochim}},
  \bibinfo{author}{\bibfnamefont{M.}~\bibnamefont{Bartenstein}},
  \bibinfo{author}{\bibfnamefont{A.}~\bibnamefont{Altmeyer}},
  \bibinfo{author}{\bibfnamefont{G.}~\bibnamefont{Hendl}},
  \bibinfo{author}{\bibfnamefont{S.}~\bibnamefont{Riedl}},
  \bibinfo{author}{\bibfnamefont{C.}~\bibnamefont{Chin}},
  \bibinfo{author}{\bibfnamefont{J.~H.} \bibnamefont{Denschlag}},
  \bibnamefont{and} \bibinfo{author}{\bibfnamefont{R.}~\bibnamefont{Grimm}},
  \bibinfo{journal}{Science} \textbf{\bibinfo{volume}{302}},
  \bibinfo{pages}{2101} (\bibinfo{year}{2003}{\natexlab{b}}).

\bibitem[{\citenamefont{Greiner et~al.}(2003)\citenamefont{Greiner, Regal, and
  Jin}}]{Greiner2003}
\bibinfo{author}{\bibfnamefont{M.}~\bibnamefont{Greiner}},
  \bibinfo{author}{\bibfnamefont{C.~A.} \bibnamefont{Regal}}, \bibnamefont{and}
  \bibinfo{author}{\bibfnamefont{D.~S.} \bibnamefont{Jin}},
  \bibinfo{journal}{Nature} \textbf{\bibinfo{volume}{426}},
  \bibinfo{pages}{537} (\bibinfo{year}{2003}).

\bibitem[{\citenamefont{Zwierlein et~al.}(2003)\citenamefont{Zwierlein, Stan,
  Schunck, Raupach, Gupta, Hadzibabic, and Ketterle}}]{Zwierlein2003}
\bibinfo{author}{\bibfnamefont{M.~W.} \bibnamefont{Zwierlein}},
  \bibinfo{author}{\bibfnamefont{C.~A.} \bibnamefont{Stan}},
  \bibinfo{author}{\bibfnamefont{C.~H.} \bibnamefont{Schunck}},
  \bibinfo{author}{\bibfnamefont{S.~M.~F.} \bibnamefont{Raupach}},
  \bibinfo{author}{\bibfnamefont{S.}~\bibnamefont{Gupta}},
  \bibinfo{author}{\bibfnamefont{Z.}~\bibnamefont{Hadzibabic}},
  \bibnamefont{and} \bibinfo{author}{\bibfnamefont{W.}~\bibnamefont{Ketterle}},
  \bibinfo{journal}{Phys.\,Rev.\,Lett.} \textbf{\bibinfo{volume}{91}},
  \bibinfo{pages}{250401} (\bibinfo{year}{2003}).

\bibitem[{\citenamefont{Inguscio et~al.}(2007)\citenamefont{Inguscio, Ketterle,
  and Salomon}}]{Inguscio2006ufg}
\bibinfo{editor}{\bibfnamefont{M.}~\bibnamefont{Inguscio}},
  \bibinfo{editor}{\bibfnamefont{W.}~\bibnamefont{Ketterle}}, \bibnamefont{and}
  \bibinfo{editor}{\bibfnamefont{C.}~\bibnamefont{Salomon}}, eds.,
  \emph{\bibinfo{title}{Ultracold Fermi Gases}} (\bibinfo{publisher}{IOS Press,
  Amsterdam}, \bibinfo{year}{2007}), \bibinfo{note}{{P}roceedings of the
  International School of Physics ``Enrico Fermi'', Course CLXIV, Varenna,
  20-30 June 2006}.

\bibitem[{\citenamefont{Chin et~al.}(2005)\citenamefont{Chin, Kraemer, Mark,
  Herbig, Waldburger, N\"{a}gerl, and Grimm}}]{Chin2005}
\bibinfo{author}{\bibfnamefont{C.}~\bibnamefont{Chin}},
  \bibinfo{author}{\bibfnamefont{T.}~\bibnamefont{Kraemer}},
  \bibinfo{author}{\bibfnamefont{M.}~\bibnamefont{Mark}},
  \bibinfo{author}{\bibfnamefont{J.}~\bibnamefont{Herbig}},
  \bibinfo{author}{\bibfnamefont{P.}~\bibnamefont{Waldburger}},
  \bibinfo{author}{\bibfnamefont{H.-C.} \bibnamefont{N\"{a}gerl}},
  \bibnamefont{and} \bibinfo{author}{\bibfnamefont{R.}~\bibnamefont{Grimm}},
  \bibinfo{journal}{Phys.\ Rev.\ Lett.} \textbf{\bibinfo{volume}{94}},
  \bibinfo{pages}{123201} (\bibinfo{year}{2005}).

\bibitem[{\citenamefont{Kraemer et~al.}(2006)\citenamefont{Kraemer, Mark,
  Waldburger, Danzl, Chin, Engeser, Lange, Pilch, Jaakkola, N\"{a}gerl
  et~al.}}]{Kraemer2006efe}
\bibinfo{author}{\bibfnamefont{T.}~\bibnamefont{Kraemer}},
  \bibinfo{author}{\bibfnamefont{M.}~\bibnamefont{Mark}},
  \bibinfo{author}{\bibfnamefont{P.}~\bibnamefont{Waldburger}},
  \bibinfo{author}{\bibfnamefont{J.~G.} \bibnamefont{Danzl}},
  \bibinfo{author}{\bibfnamefont{C.}~\bibnamefont{Chin}},
  \bibinfo{author}{\bibfnamefont{B.}~\bibnamefont{Engeser}},
  \bibinfo{author}{\bibfnamefont{A.~D.} \bibnamefont{Lange}},
  \bibinfo{author}{\bibfnamefont{K.}~\bibnamefont{Pilch}},
  \bibinfo{author}{\bibfnamefont{A.}~\bibnamefont{Jaakkola}},
  \bibinfo{author}{\bibfnamefont{H.-C.} \bibnamefont{N\"{a}gerl}},
  \bibnamefont{et~al.}, \bibinfo{journal}{Nature}
  \textbf{\bibinfo{volume}{440}}, \bibinfo{pages}{315} (\bibinfo{year}{2006}).

\bibitem[{\citenamefont{St\"oferle et~al.}(2006)\citenamefont{St\"oferle,
  Moritz, G\"unter, K\"ohl, and Esslinger}}]{Stoferle2006mof}
\bibinfo{author}{\bibfnamefont{T.}~\bibnamefont{St\"oferle}},
  \bibinfo{author}{\bibfnamefont{H.}~\bibnamefont{Moritz}},
  \bibinfo{author}{\bibfnamefont{K.}~\bibnamefont{G\"unter}},
  \bibinfo{author}{\bibfnamefont{M.}~\bibnamefont{K\"ohl}}, \bibnamefont{and}
  \bibinfo{author}{\bibfnamefont{T.}~\bibnamefont{Esslinger}},
  \bibinfo{journal}{Phys. Rev. Lett.} \textbf{\bibinfo{volume}{96}},
  \bibinfo{eid}{030401} (\bibinfo{year}{2006}).

\bibitem[{\citenamefont{Thalhammer et~al.}(2006)\citenamefont{Thalhammer,
  Winkler, Lang, Schmid, Grimm, and {Hecker Denschlag}}}]{Thalhammer2006llf}
\bibinfo{author}{\bibfnamefont{G.}~\bibnamefont{Thalhammer}},
  \bibinfo{author}{\bibfnamefont{K.}~\bibnamefont{Winkler}},
  \bibinfo{author}{\bibfnamefont{F.}~\bibnamefont{Lang}},
  \bibinfo{author}{\bibfnamefont{S.}~\bibnamefont{Schmid}},
  \bibinfo{author}{\bibfnamefont{R.}~\bibnamefont{Grimm}}, \bibnamefont{and}
  \bibinfo{author}{\bibfnamefont{J.}~\bibnamefont{{Hecker Denschlag}}},
  \bibinfo{journal}{Phys. Rev. Lett.} \textbf{\bibinfo{volume}{96}},
  \bibinfo{eid}{050402} (\bibinfo{year}{2006}).

\bibitem[{\citenamefont{Ospelkaus et~al.}(2006)\citenamefont{Ospelkaus,
  Ospelkaus, Humbert, Ernst, Sengstock, and Bongs}}]{Ospelkaus2006uhm}
\bibinfo{author}{\bibfnamefont{C.}~\bibnamefont{Ospelkaus}},
  \bibinfo{author}{\bibfnamefont{S.}~\bibnamefont{Ospelkaus}},
  \bibinfo{author}{\bibfnamefont{L.}~\bibnamefont{Humbert}},
  \bibinfo{author}{\bibfnamefont{P.}~\bibnamefont{Ernst}},
  \bibinfo{author}{\bibfnamefont{K.}~\bibnamefont{Sengstock}},
  \bibnamefont{and} \bibinfo{author}{\bibfnamefont{K.}~\bibnamefont{Bongs}},
  \bibinfo{journal}{Phys. Rev. Lett.} \textbf{\bibinfo{volume}{97}},
  \bibinfo{eid}{120402} (\bibinfo{year}{2006}).

\bibitem[{\citenamefont{Winkler et~al.}(2006)\citenamefont{Winkler, Thalhammer,
  Lang, Grimm, Hecker~Denschlag, Daley, Kantian, B{\"u}chler, and
  Zoller}}]{Winkler2006rba}
\bibinfo{author}{\bibfnamefont{K.}~\bibnamefont{Winkler}},
  \bibinfo{author}{\bibfnamefont{G.}~\bibnamefont{Thalhammer}},
  \bibinfo{author}{\bibfnamefont{F.}~\bibnamefont{Lang}},
  \bibinfo{author}{\bibfnamefont{R.}~\bibnamefont{Grimm}},
  \bibinfo{author}{\bibfnamefont{J.}~\bibnamefont{Hecker~Denschlag}},
  \bibinfo{author}{\bibfnamefont{A.~J.} \bibnamefont{Daley}},
  \bibinfo{author}{\bibfnamefont{A.}~\bibnamefont{Kantian}},
  \bibinfo{author}{\bibfnamefont{H.~P.} \bibnamefont{B{\"u}chler}},
  \bibnamefont{and} \bibinfo{author}{\bibfnamefont{P.}~\bibnamefont{Zoller}},
  \bibinfo{journal}{Nature} \textbf{\bibinfo{volume}{441}},
  \bibinfo{pages}{853} (\bibinfo{year}{2006}).

\bibitem[{\citenamefont{Volz et~al.}(2006)\citenamefont{Volz, Syassen, Bauer,
  Hansis, D{\"u}rr, and Rempe}}]{Volz2006pqs}
\bibinfo{author}{\bibfnamefont{T.}~\bibnamefont{Volz}},
  \bibinfo{author}{\bibfnamefont{N.}~\bibnamefont{Syassen}},
  \bibinfo{author}{\bibfnamefont{D.}~\bibnamefont{Bauer}},
  \bibinfo{author}{\bibfnamefont{E.}~\bibnamefont{Hansis}},
  \bibinfo{author}{\bibfnamefont{S.}~\bibnamefont{D{\"u}rr}}, \bibnamefont{and}
  \bibinfo{author}{\bibfnamefont{G.}~\bibnamefont{Rempe}},
  \bibinfo{journal}{Nature Phys.} \textbf{\bibinfo{volume}{2}},
  \bibinfo{pages}{692} (\bibinfo{year}{2006}).

\bibitem[{\citenamefont{Tiesinga et~al.}(1993)\citenamefont{Tiesinga, Verhaar,
  and Stoof}}]{Tiesinga1993}
\bibinfo{author}{\bibfnamefont{E.}~\bibnamefont{Tiesinga}},
  \bibinfo{author}{\bibfnamefont{B.~J.} \bibnamefont{Verhaar}},
  \bibnamefont{and} \bibinfo{author}{\bibfnamefont{H.~T.~C.}
  \bibnamefont{Stoof}}, \bibinfo{journal}{Phys.\,Rev.\,A}
  \textbf{\bibinfo{volume}{47}}, \bibinfo{pages}{4114} (\bibinfo{year}{1993}).

\bibitem[{\citenamefont{Inouye et~al.}(1998)\citenamefont{Inouye, Andrews,
  Stenger, Miesner, Stamper-Kurn, and Ketterle}}]{Inouye1998}
\bibinfo{author}{\bibfnamefont{S.}~\bibnamefont{Inouye}},
  \bibinfo{author}{\bibfnamefont{M.~R.} \bibnamefont{Andrews}},
  \bibinfo{author}{\bibfnamefont{J.}~\bibnamefont{Stenger}},
  \bibinfo{author}{\bibfnamefont{H.-J.} \bibnamefont{Miesner}},
  \bibinfo{author}{\bibfnamefont{S.~M.} \bibnamefont{Stamper-Kurn}},
  \bibnamefont{and} \bibinfo{author}{\bibfnamefont{W.}~\bibnamefont{Ketterle}},
  \bibinfo{journal}{Nature} \textbf{\bibinfo{volume}{392}},
  \bibinfo{pages}{151} (\bibinfo{year}{1998}).

\bibitem[{\citenamefont{Chin et~al.}(2004)\citenamefont{Chin, Vuletic, Kerman,
  Chu, Tiesinga, Leo, and Williams}}]{Chin2004}
\bibinfo{author}{\bibfnamefont{C.}~\bibnamefont{Chin}},
  \bibinfo{author}{\bibfnamefont{V.}~\bibnamefont{Vuletic}},
  \bibinfo{author}{\bibfnamefont{A.~J.} \bibnamefont{Kerman}},
  \bibinfo{author}{\bibfnamefont{S.}~\bibnamefont{Chu}},
  \bibinfo{author}{\bibfnamefont{E.}~\bibnamefont{Tiesinga}},
  \bibinfo{author}{\bibfnamefont{P.~J.} \bibnamefont{Leo}}, \bibnamefont{and}
  \bibinfo{author}{\bibfnamefont{C.~J.} \bibnamefont{Williams}},
  \bibinfo{journal}{Phys.\ Rev.\ A} \textbf{\bibinfo{volume}{70}},
  \bibinfo{pages}{032701} (\bibinfo{year}{2004}).

\bibitem[{\citenamefont{Weber et~al.}(2003{\natexlab{a}})\citenamefont{Weber,
  Herbig, Mark, N\"{a}gerl, and Grimm}}]{Weber2003}
\bibinfo{author}{\bibfnamefont{T.}~\bibnamefont{Weber}},
  \bibinfo{author}{\bibfnamefont{J.}~\bibnamefont{Herbig}},
  \bibinfo{author}{\bibfnamefont{M.}~\bibnamefont{Mark}},
  \bibinfo{author}{\bibfnamefont{H.-C.} \bibnamefont{N\"{a}gerl}},
  \bibnamefont{and} \bibinfo{author}{\bibfnamefont{R.}~\bibnamefont{Grimm}},
  \bibinfo{journal}{Science} \textbf{\bibinfo{volume}{299}},
  \bibinfo{pages}{232} (\bibinfo{year}{2003}{\natexlab{a}}).

\bibitem[{\citenamefont{Vuleti\'{c} et~al.}(1999)\citenamefont{Vuleti\'{c},
  Kerman, Chin, and Chu}}]{Vuletic1999}
\bibinfo{author}{\bibfnamefont{V.}~\bibnamefont{Vuleti\'{c}}},
  \bibinfo{author}{\bibfnamefont{A.~J.} \bibnamefont{Kerman}},
  \bibinfo{author}{\bibfnamefont{C.}~\bibnamefont{Chin}}, \bibnamefont{and}
  \bibinfo{author}{\bibfnamefont{S.}~\bibnamefont{Chu}},
  \bibinfo{journal}{Phys.\ Rev.\ Lett.} \textbf{\bibinfo{volume}{82}},
  \bibinfo{pages}{1406} (\bibinfo{year}{1999}).

\bibitem[{\citenamefont{Chin et~al.}(2000)\citenamefont{Chin, Vuleti\'{c},
  Kerman, and Chu}}]{Chin2000}
\bibinfo{author}{\bibfnamefont{C.}~\bibnamefont{Chin}},
  \bibinfo{author}{\bibfnamefont{V.}~\bibnamefont{Vuleti\'{c}}},
  \bibinfo{author}{\bibfnamefont{A.~J.} \bibnamefont{Kerman}},
  \bibnamefont{and} \bibinfo{author}{\bibfnamefont{S.}~\bibnamefont{Chu}},
  \bibinfo{journal}{Phys.\ Rev.\ Lett.} \textbf{\bibinfo{volume}{85}},
  \bibinfo{pages}{2717} (\bibinfo{year}{2000}).

\bibitem[{\citenamefont{Leo et~al.}(2000)\citenamefont{Leo, Williams, and
  Julienne}}]{Leo2000}
\bibinfo{author}{\bibfnamefont{P.~J.} \bibnamefont{Leo}},
  \bibinfo{author}{\bibfnamefont{C.~J.} \bibnamefont{Williams}},
  \bibnamefont{and} \bibinfo{author}{\bibfnamefont{P.~S.}
  \bibnamefont{Julienne}}, \bibinfo{journal}{Phys.\ Rev.\ Lett.}
  \textbf{\bibinfo{volume}{85}}, \bibinfo{pages}{2721} (\bibinfo{year}{2000}).

\bibitem[{\citenamefont{Takekoshi et~al.}(1998)\citenamefont{Takekoshi,
  Patterson, and Knize}}]{Takekoshi1998}
\bibinfo{author}{\bibfnamefont{T.}~\bibnamefont{Takekoshi}},
  \bibinfo{author}{\bibfnamefont{B.~M.} \bibnamefont{Patterson}},
  \bibnamefont{and} \bibinfo{author}{\bibfnamefont{R.~J.} \bibnamefont{Knize}},
  \bibinfo{journal}{Phys.\, Rev.\, Lett.\,} \textbf{\bibinfo{volume}{81}},
  \bibinfo{pages}{5105} (\bibinfo{year}{1998}).

\bibitem[{\citenamefont{Staanum et~al.}(2006)\citenamefont{Staanum, Kraft,
  Lange, Wester, and Weidem{\"u}ller}}]{Staanum2006eio}
\bibinfo{author}{\bibfnamefont{P.}~\bibnamefont{Staanum}},
  \bibinfo{author}{\bibfnamefont{S.~D.} \bibnamefont{Kraft}},
  \bibinfo{author}{\bibfnamefont{J.}~\bibnamefont{Lange}},
  \bibinfo{author}{\bibfnamefont{R.}~\bibnamefont{Wester}}, \bibnamefont{and}
  \bibinfo{author}{\bibfnamefont{M.}~\bibnamefont{Weidem{\"u}ller}},
  \bibinfo{journal}{Phys. Rev. Lett.} \textbf{\bibinfo{volume}{96}},
  \bibinfo{eid}{023201} (\bibinfo{year}{2006}).

\bibitem[{\citenamefont{Zahzam et~al.}(2006)\citenamefont{Zahzam, Vogt,
  Mudrich, Comparat, and Pillet}}]{Zahzam2006amc}
\bibinfo{author}{\bibfnamefont{N.}~\bibnamefont{Zahzam}},
  \bibinfo{author}{\bibfnamefont{T.}~\bibnamefont{Vogt}},
  \bibinfo{author}{\bibfnamefont{M.}~\bibnamefont{Mudrich}},
  \bibinfo{author}{\bibfnamefont{D.}~\bibnamefont{Comparat}}, \bibnamefont{and}
  \bibinfo{author}{\bibfnamefont{P.}~\bibnamefont{Pillet}},
  \bibinfo{journal}{Phys. Rev. Lett.} \textbf{\bibinfo{volume}{96}},
  \bibinfo{eid}{023202} (\bibinfo{year}{2006}).

\bibitem[{\citenamefont{Russell et~al.}(1929)\citenamefont{Russell, Shenstone,
  and Turner}}]{Russell1929}
\bibinfo{author}{\bibfnamefont{H.~N.} \bibnamefont{Russell}},
  \bibinfo{author}{\bibfnamefont{A.~G.} \bibnamefont{Shenstone}},
  \bibnamefont{and} \bibinfo{author}{\bibfnamefont{L.~A.}
  \bibnamefont{Turner}}, \bibinfo{journal}{Phys.\ Rev.}
  \textbf{\bibinfo{volume}{33}}, \bibinfo{pages}{900} (\bibinfo{year}{1929}).

\bibitem[{\citenamefont{Mies et~al.}(1996)\citenamefont{Mies, Williams,
  Julienne, and Krauss}}]{Mies1996}
\bibinfo{author}{\bibfnamefont{F.~H.} \bibnamefont{Mies}},
  \bibinfo{author}{\bibfnamefont{C.~J.} \bibnamefont{Williams}},
  \bibinfo{author}{\bibfnamefont{P.~S.} \bibnamefont{Julienne}},
  \bibnamefont{and} \bibinfo{author}{\bibfnamefont{M.}~\bibnamefont{Krauss}},
  \bibinfo{journal}{J.\ Res.\ Natl.\ Inst.\ Stan.}
  \textbf{\bibinfo{volume}{101}}, \bibinfo{pages}{521} (\bibinfo{year}{1996}).

\bibitem[{\citenamefont{Gao}(2000)}]{Gao2000}
\bibinfo{author}{\bibfnamefont{B.}~\bibnamefont{Gao}}, \bibinfo{journal}{Phys.\
  Rev.\ A} \textbf{\bibinfo{volume}{62}}, \bibinfo{pages}{050702}
  (\bibinfo{year}{2000}).

\bibitem[{\citenamefont{Tiesinga and Julienne}(2007)}]{TiesingaPrivate}
\bibinfo{author}{\bibfnamefont{E.}~\bibnamefont{Tiesinga}} \bibnamefont{and}
  \bibinfo{author}{\bibfnamefont{P.~S.} \bibnamefont{Julienne}},
  \bibinfo{journal}{private communication}  (\bibinfo{year}{2007}).

\bibitem[{\citenamefont{Kraemer et~al.}(2004)\citenamefont{Kraemer, Herbig,
  Mark, Weber, Chin, N\"{a}gerl, and Grimm}}]{Kraemer2004}
\bibinfo{author}{\bibfnamefont{T.}~\bibnamefont{Kraemer}},
  \bibinfo{author}{\bibfnamefont{J.}~\bibnamefont{Herbig}},
  \bibinfo{author}{\bibfnamefont{M.}~\bibnamefont{Mark}},
  \bibinfo{author}{\bibfnamefont{T.}~\bibnamefont{Weber}},
  \bibinfo{author}{\bibfnamefont{C.}~\bibnamefont{Chin}},
  \bibinfo{author}{\bibfnamefont{H.-C.} \bibnamefont{N\"{a}gerl}},
  \bibnamefont{and} \bibinfo{author}{\bibfnamefont{R.}~\bibnamefont{Grimm}},
  \bibinfo{journal}{Appl.\, Phys.\, B} \textbf{\bibinfo{volume}{79}},
  \bibinfo{pages}{1013} (\bibinfo{year}{2004}).

\bibitem[{\citenamefont{Treutlein et~al.}(2001)\citenamefont{Treutlein, Chung,
  and Chu}}]{Treutlein2001}
\bibinfo{author}{\bibfnamefont{P.}~\bibnamefont{Treutlein}},
  \bibinfo{author}{\bibfnamefont{K.~Y.} \bibnamefont{Chung}}, \bibnamefont{and}
  \bibinfo{author}{\bibfnamefont{S.}~\bibnamefont{Chu}},
  \bibinfo{journal}{Phys.\, Rev.\, A} \textbf{\bibinfo{volume}{63}},
  \bibinfo{pages}{051401} (\bibinfo{year}{2001}).

\bibitem[{\citenamefont{Weber et~al.}(2003{\natexlab{b}})\citenamefont{Weber,
  Herbig, Mark, N\"{a}gerl, and Grimm}}]{Weber2003a}
\bibinfo{author}{\bibfnamefont{T.}~\bibnamefont{Weber}},
  \bibinfo{author}{\bibfnamefont{J.}~\bibnamefont{Herbig}},
  \bibinfo{author}{\bibfnamefont{M.}~\bibnamefont{Mark}},
  \bibinfo{author}{\bibfnamefont{H.-C.} \bibnamefont{N\"{a}gerl}},
  \bibnamefont{and} \bibinfo{author}{\bibfnamefont{R.}~\bibnamefont{Grimm}},
  \bibinfo{journal}{Phys.\, Rev.\, Lett.\,} \textbf{\bibinfo{volume}{91}},
  \bibinfo{pages}{123201} (\bibinfo{year}{2003}{\natexlab{b}}).

\bibitem[{\citenamefont{Mark et~al.}(2005)\citenamefont{Mark, Kraemer, Herbig,
  Chin, N\"{a}gerl, and Grimm}}]{Mark2005}
\bibinfo{author}{\bibfnamefont{M.}~\bibnamefont{Mark}},
  \bibinfo{author}{\bibfnamefont{T.}~\bibnamefont{Kraemer}},
  \bibinfo{author}{\bibfnamefont{J.}~\bibnamefont{Herbig}},
  \bibinfo{author}{\bibfnamefont{C.}~\bibnamefont{Chin}},
  \bibinfo{author}{\bibfnamefont{H.-C.} \bibnamefont{N\"{a}gerl}},
  \bibnamefont{and} \bibinfo{author}{\bibfnamefont{R.}~\bibnamefont{Grimm}},
  \bibinfo{journal}{Europhys.\, Lett.\,} \textbf{\bibinfo{volume}{69}},
  \bibinfo{pages}{706} (\bibinfo{year}{2005}).

\bibitem[{\citenamefont{Mukaiyama et~al.}(2004)\citenamefont{Mukaiyama,
  Abo-Shaeer, Xu, Chin, and Ketterle}}]{Mukaiyama2004}
\bibinfo{author}{\bibfnamefont{T.}~\bibnamefont{Mukaiyama}},
  \bibinfo{author}{\bibfnamefont{J.~R.} \bibnamefont{Abo-Shaeer}},
  \bibinfo{author}{\bibfnamefont{K.}~\bibnamefont{Xu}},
  \bibinfo{author}{\bibfnamefont{J.~K.} \bibnamefont{Chin}}, \bibnamefont{and}
  \bibinfo{author}{\bibfnamefont{W.}~\bibnamefont{Ketterle}},
  \bibinfo{journal}{Phys.\,Rev.\,Lett.} \textbf{\bibinfo{volume}{92}},
  \bibinfo{pages}{180402} (\bibinfo{year}{2004}).

\bibitem[{\citenamefont{Mark et~al.}(2007)\citenamefont{Mark, Kraemer,
  Waldburger, Herbig, Chin, N\"agerl, and Grimm}}]{Mark2007}
\bibinfo{author}{\bibfnamefont{M.}~\bibnamefont{Mark}},
  \bibinfo{author}{\bibfnamefont{T.}~\bibnamefont{Kraemer}},
  \bibinfo{author}{\bibfnamefont{P.}~\bibnamefont{Waldburger}},
  \bibinfo{author}{\bibfnamefont{J.}~\bibnamefont{Herbig}},
  \bibinfo{author}{\bibfnamefont{C.}~\bibnamefont{Chin}},
  \bibinfo{author}{\bibfnamefont{H.-C.} \bibnamefont{N\"agerl}},
  \bibnamefont{and} \bibinfo{author}{\bibfnamefont{R.}~\bibnamefont{Grimm}},
  \bibinfo{journal}{subm. for publication, arXiv:0704.0653}
  (\bibinfo{year}{2007}).

\bibitem[{\citenamefont{Landau}(1932)}]{Landau1932xxx}
\bibinfo{author}{\bibfnamefont{L.}~\bibnamefont{Landau}},
  \bibinfo{journal}{Phys. Z. Sowjetunion} \textbf{\bibinfo{volume}{2}},
  \bibinfo{pages}{46} (\bibinfo{year}{1932}).

\bibitem[{\citenamefont{Zener}(1932)}]{Zener1932xxx}
\bibinfo{author}{\bibfnamefont{C.}~\bibnamefont{Zener}},
  \bibinfo{journal}{Proc. R. Soc. London, Ser. A}
  \textbf{\bibinfo{volume}{137}}, \bibinfo{pages}{696} (\bibinfo{year}{1932}).

\bibitem[{\citenamefont{D\"{u}rr
  et~al.}(2004{\natexlab{b}})\citenamefont{D\"{u}rr, Volz, and
  Rempe}}]{Duerr2004}
\bibinfo{author}{\bibfnamefont{S.}~\bibnamefont{D\"{u}rr}},
  \bibinfo{author}{\bibfnamefont{T.}~\bibnamefont{Volz}}, \bibnamefont{and}
  \bibinfo{author}{\bibfnamefont{G.}~\bibnamefont{Rempe}},
  \bibinfo{journal}{Phys.\, Rev.\, A} \textbf{\bibinfo{volume}{70}},
  \bibinfo{pages}{031601(R)} (\bibinfo{year}{2004}{\natexlab{b}}).

\bibitem[{\citenamefont{Knoop et~al.}(2007)\citenamefont{Knoop, Mark, Ferlaino,
  Danzl, Kraemer, N\"{a}gerl, and Grimm}}]{Knoop2007}
\bibinfo{author}{\bibfnamefont{S.}~\bibnamefont{Knoop}},
  \bibinfo{author}{\bibfnamefont{M.}~\bibnamefont{Mark}},
  \bibinfo{author}{\bibfnamefont{F.}~\bibnamefont{Ferlaino}},
  \bibinfo{author}{\bibfnamefont{J.~G.} \bibnamefont{Danzl}},
  \bibinfo{author}{\bibfnamefont{T.}~\bibnamefont{Kraemer}},
  \bibinfo{author}{\bibfnamefont{H.-C.} \bibnamefont{N\"{a}gerl}},
  \bibnamefont{and} \bibinfo{author}{\bibfnamefont{R.}~\bibnamefont{Grimm}},
  \bibinfo{journal}{manuscript in preparation}  (\bibinfo{year}{2007}).

\bibitem[{\citenamefont{Claussen et~al.}(2003)\citenamefont{Claussen,
  Kokkelmans, Thompson, Donley, Hodby, and Wieman}}]{Claussen2003}
\bibinfo{author}{\bibfnamefont{N.~R.} \bibnamefont{Claussen}},
  \bibinfo{author}{\bibfnamefont{S.~J.~J.~M.~F.} \bibnamefont{Kokkelmans}},
  \bibinfo{author}{\bibfnamefont{S.~T.} \bibnamefont{Thompson}},
  \bibinfo{author}{\bibfnamefont{E.~A.} \bibnamefont{Donley}},
  \bibinfo{author}{\bibfnamefont{E.}~\bibnamefont{Hodby}}, \bibnamefont{and}
  \bibinfo{author}{\bibfnamefont{C.~E.} \bibnamefont{Wieman}},
  \bibinfo{journal}{Phys.\ Rev.\ A} \textbf{\bibinfo{volume}{67}},
  \bibinfo{pages}{060701(R)} (\bibinfo{year}{2003}).

\bibitem[{\citenamefont{Thompson
  et~al.}(2005{\natexlab{a}})\citenamefont{Thompson, Hodby, and
  Wieman}}]{Thompson2005}
\bibinfo{author}{\bibfnamefont{S.~T.} \bibnamefont{Thompson}},
  \bibinfo{author}{\bibfnamefont{E.}~\bibnamefont{Hodby}}, \bibnamefont{and}
  \bibinfo{author}{\bibfnamefont{C.~E.} \bibnamefont{Wieman}},
  \bibinfo{journal}{Phys.\ Rev.\ Lett.} \textbf{\bibinfo{volume}{95}},
  \bibinfo{pages}{190404} (\bibinfo{year}{2005}{\natexlab{a}}).

\bibitem[{\citenamefont{Bartenstein et~al.}(2005)\citenamefont{Bartenstein,
  Altmeyer, Riedl, Geursen, Jochim, Chin, Denschlag, Grimm, Simoni, Tiesinga
  et~al.}}]{Bartenst2004b}
\bibinfo{author}{\bibfnamefont{M.}~\bibnamefont{Bartenstein}},
  \bibinfo{author}{\bibfnamefont{A.}~\bibnamefont{Altmeyer}},
  \bibinfo{author}{\bibfnamefont{S.}~\bibnamefont{Riedl}},
  \bibinfo{author}{\bibfnamefont{R.}~\bibnamefont{Geursen}},
  \bibinfo{author}{\bibfnamefont{S.}~\bibnamefont{Jochim}},
  \bibinfo{author}{\bibfnamefont{C.}~\bibnamefont{Chin}},
  \bibinfo{author}{\bibfnamefont{J.~H.} \bibnamefont{Denschlag}},
  \bibinfo{author}{\bibfnamefont{R.}~\bibnamefont{Grimm}},
  \bibinfo{author}{\bibfnamefont{A.}~\bibnamefont{Simoni}},
  \bibinfo{author}{\bibfnamefont{E.}~\bibnamefont{Tiesinga}},
  \bibnamefont{et~al.}, \bibinfo{journal}{Phys.\,Rev.\,Lett.}
  \textbf{\bibinfo{volume}{94}}, \bibinfo{pages}{103201}
  (\bibinfo{year}{2005}).

\bibitem[{\citenamefont{Thompson
  et~al.}(2005{\natexlab{b}})\citenamefont{Thompson, Hodby, and
  Wieman}}]{Thompson2005b}
\bibinfo{author}{\bibfnamefont{S.~T.} \bibnamefont{Thompson}},
  \bibinfo{author}{\bibfnamefont{E.}~\bibnamefont{Hodby}}, \bibnamefont{and}
  \bibinfo{author}{\bibfnamefont{C.~E.} \bibnamefont{Wieman}},
  \bibinfo{journal}{Phys.\, Rev.\, Lett.\,} \textbf{\bibinfo{volume}{94}},
  \bibinfo{pages}{020401} (\bibinfo{year}{2005}{\natexlab{b}}).

\bibitem[{\citenamefont{Julienne et~al.}(2004)\citenamefont{Julienne, Tiesinga,
  and K\"{o}hler}}]{Julienne2004}
\bibinfo{author}{\bibfnamefont{P.~S.} \bibnamefont{Julienne}},
  \bibinfo{author}{\bibfnamefont{E.}~\bibnamefont{Tiesinga}}, \bibnamefont{and}
  \bibinfo{author}{\bibfnamefont{T.}~\bibnamefont{K\"{o}hler}},
  \bibinfo{journal}{J.\,Mod.\, Opt.} \textbf{\bibinfo{volume}{513}},
  \bibinfo{pages}{1787} (\bibinfo{year}{2004}).

\bibitem[{\citenamefont{Jensen et~al.}(2004)\citenamefont{Jensen, Riisager,
  Fedorov, and Garrido}}]{Jensen2004}
\bibinfo{author}{\bibfnamefont{A.~S.} \bibnamefont{Jensen}},
  \bibinfo{author}{\bibfnamefont{K.}~\bibnamefont{Riisager}},
  \bibinfo{author}{\bibfnamefont{D.~V.} \bibnamefont{Fedorov}},
  \bibnamefont{and} \bibinfo{author}{\bibfnamefont{E.}~\bibnamefont{Garrido}},
  \bibinfo{journal}{Rev.\, Mod.\, Phys.\,} \textbf{\bibinfo{volume}{76}},
  \bibinfo{pages}{215} (\bibinfo{year}{2004}).

\bibitem[{\citenamefont{K\"{o}hler et~al.}(2005)\citenamefont{K\"{o}hler,
  Tiesinga, and Julienne}}]{Koehler2005}
\bibinfo{author}{\bibfnamefont{T.}~\bibnamefont{K\"{o}hler}},
  \bibinfo{author}{\bibfnamefont{E.}~\bibnamefont{Tiesinga}}, \bibnamefont{and}
  \bibinfo{author}{\bibfnamefont{P.~S.} \bibnamefont{Julienne}},
  \bibinfo{journal}{Phys.\, Rev.\, Lett.\,} \textbf{\bibinfo{volume}{94}},
  \bibinfo{pages}{020402} (\bibinfo{year}{2005}).

\bibitem[{\citenamefont{Chin and Julienne}(2005)}]{Chin2005PRA}
\bibinfo{author}{\bibfnamefont{C.}~\bibnamefont{Chin}} \bibnamefont{and}
  \bibinfo{author}{\bibfnamefont{P.~S.} \bibnamefont{Julienne}},
  \bibinfo{journal}{Phys. Rev. A} \textbf{\bibinfo{volume}{71}},
  \bibinfo{pages}{012713} (\bibinfo{year}{2005}).

\bibitem[{\citenamefont{Weber}(2003)}]{Weber2003PhD}
\bibinfo{author}{\bibfnamefont{T.}~\bibnamefont{Weber}}, \bibinfo{type}{Phd
  thesis}, \bibinfo{school}{University of Innsbruck} (\bibinfo{year}{2003}),
  \bibinfo{note}{downloadable at: www.ultracold.at/theses}.

\end{thebibliography}

\end{document}